\PassOptionsToPackage{numbers,sort&compress}{natbib}
\documentclass{article} % For LaTeX2e
\usepackage[preprint]{neurips_2024}
\usepackage[export]{adjustbox}

\usepackage{amsmath,amsfonts,bm}

\def\eqref#1{equation~\ref{#1}}
\def\1{\bm{1}}

\DeclareMathAlphabet{\mathsfit}{\encodingdefault}{\sfdefault}{m}{sl}
\SetMathAlphabet{\mathsfit}{bold}{\encodingdefault}{\sfdefault}{bx}{n}

\usepackage{wrapfig}
\usepackage{authblk}
\usepackage{wrapfig}
\usepackage{array}
\usepackage{booktabs}
\usepackage{amssymb}
\usepackage[utf8]{inputenc} % allow utf-8 input
\usepackage[T1]{fontenc}    % use 8-bit T1 fonts
\usepackage{url}            % simple URL typesetting
\usepackage{booktabs}       % professional-quality tables
\usepackage{amsfonts}       % blackboard math symbols
\usepackage{nicefrac}       % compact symbols for 1/2, etc.
\usepackage{microtype}      % microtypography
\usepackage{xcolor}         % colors
\usepackage{amsmath}
\usepackage{booktabs}
\usepackage{multicol}
\usepackage{multirow}
\usepackage{graphicx}
\usepackage{float}
\usepackage{enumitem}
\usepackage{xfrac}
\usepackage{svg}
\usepackage{xcolor}
\usepackage{colortbl}
\usepackage{subcaption}
\usepackage{pifont}
\usepackage{hyperref}
\hypersetup{
    colorlinks,
    citecolor=[rgb]{0.00, 0.45, 0.70},
    linkcolor=[rgb]{0.01, 0.62, 0.45},
    urlcolor=[rgb]{0.80, 0.47, 0.74},
    }

\usepackage{colortbl}

\title{Exact Aggregation for Federated and Efficient Fine-Tuning of Foundation Models}

\author[1]{\textbf{Raghav Singhal}\textsuperscript{\textbf{$\boldsymbol{\ddag}$}}}
\author[1]{\textbf{Kaustubh Ponkshe}\textsuperscript{\textbf{$\boldsymbol{\ddag}$}}}
\author[1, 2]{\textbf{Praneeth Vepakomma}}
\affil[1]{Mohamed bin Zayed University of Artificial Intelligence, UAE}
\affil[2]{Massachusetts Institute of Technology, USA}

\begin{document}

\maketitle
\def\thefootnote{\textbf{$\boldsymbol{\ddag}$}}\footnotetext{Equal contributions. Author ordering determined by coin flip over a Google Meet.}\def\thefootnote{\arabic{footnote}}

\begin{abstract}
Low-Rank Adaptation (LoRA) is a popular technique for efficient fine-tuning of foundation models. However, applying LoRA in federated learning environments, where data is distributed across multiple clients, presents unique challenges. Existing methods rely on traditional federated averaging of LoRA adapters, resulting in inexact updates. To address this, we propose \textbf{Fed}erated \textbf{Ex}act \textbf{LoRA}, or \textbf{FedEx-LoRA}, which adds a residual error term to the pretrained frozen weight matrix. Our approach achieves exact updates with minimal computational and communication overhead, preserving LoRA’s efficiency. We evaluate the method on various models across arithmetic reasoning, commonsense reasoning, natural language understanding and natural language generation tasks, showing consistent performance gains over state-of-the-art methods across multiple settings. Through extensive analysis, we quantify that the deviations in updates from the ideal solution are significant, highlighting the need for exact aggregation. Our method's simplicity, efficiency, and broad applicability position it as a promising solution for accurate and effective federated fine-tuning of foundation models. We have released our code publicly at \url{https://github.com/RaghavSinghal10/fedex-lora}.
\end{abstract}

\section{Introduction}
\label{intro}

The introduction of large language models (LLMs) has revolutionized natural language processing, enabling unprecedented performance across a wide range of tasks \citep{achiam2023gpt, touvron2023llama-2, team2023gemini, chang2024survey, raffel2020exploring, zeng2022glm}. While these models excel at transfer learning, their true potential is often unlocked through fine-tuning — a critical process that aligns these general-purpose models with specific tasks or domains. Moreover, the sheer size of these models presents significant challenges for fine-tuning and deployment, particularly in resource-constrained or distributed environments. To address these challenges, parameter-efficient fine-tuning (PEFT) methods have gained prominence, with Low-Rank Adaptation (LoRA) emerging as a particularly effective approach \citep{lora}. LoRA's success lies in its ability to adapt LLMs to new tasks by training only a small number of parameters, while freezing rest of the parameters. This significantly reduces computational and memory requirements without compromising performance. Although good progress in training of LLMs has been realized by entities equipped with massive computational resources, there is hoards of unreachable data in verticals such as healthcare, finance, law firms, social-media and logistics. Federated learning (FL) is a popular paradigm to learn a machine learning model in this setting with multiple distributed entities \citep{konečný2017federatedlearningstrategiesimproving, kairouz2021advances,bonawitz2019federatedlearningscaledesign} holding siloed data. 

%However, a vanilla application of LoRA in federated settings introduces new challenges that lead to suboptimal aggregation of LoRA adapters, potentially degrading model performance and convergence rates. 

Federated Fine-Tuning (FFT) for foundation models addresses the challenge of leveraging distributed datasets while preserving data privacy. %Early approaches adapted Parameter-Efficient Fine-Tuning (PEFT) methods like Low-Rank Adaptation (LoRA) to federated settings. 
 The current state-of-the-art, Federated Instruction Tuning (FedIT, \citet{FedIT}), uses conventional federated aggregation to average the low-rank matrices $\mathbf{A}$ and $\mathbf{B}$ individually. The resulting update matrix which is formed post aggregation is thus the product of the averaged matrices $\mathbf{A}$  and $\mathbf{B}$. However, the ideal update should be the average of the products of the low-rank adapters $\mathbf{A}$ and $\mathbf{B}$. The discrepancy results from the fact that \textit{"the average of the products is not equal to the product of the averages"}.
A naive adhoc intervention of modifying the aggregation to directly average the client updates is not a viable solution, since the subsequently obtained weight matrix loses its low-rank structure. The low-rank structure provides the efficiency benefits of LoRA in the first place, making this approach computationally intractable.

\begin{figure}[ht]
    \centering
    % First subfigure
    \begin{subfigure}{0.47\textwidth}
        \centering
        \includegraphics[width=\textwidth]{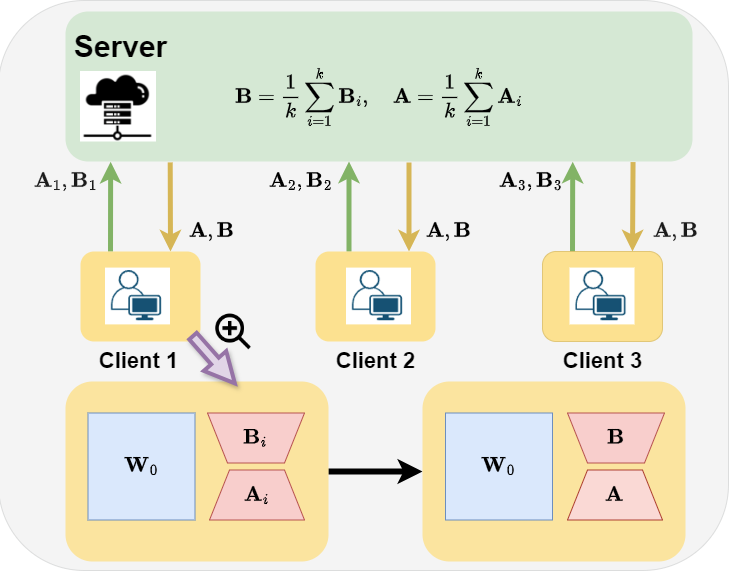}
        \caption{FedIT}
        \label{fig:rounds-a0-main}
    \end{subfigure}
    %\hfill
    % Second subfigure
    \begin{subfigure}{0.47\textwidth}
        \centering
        \includegraphics[width=\textwidth]{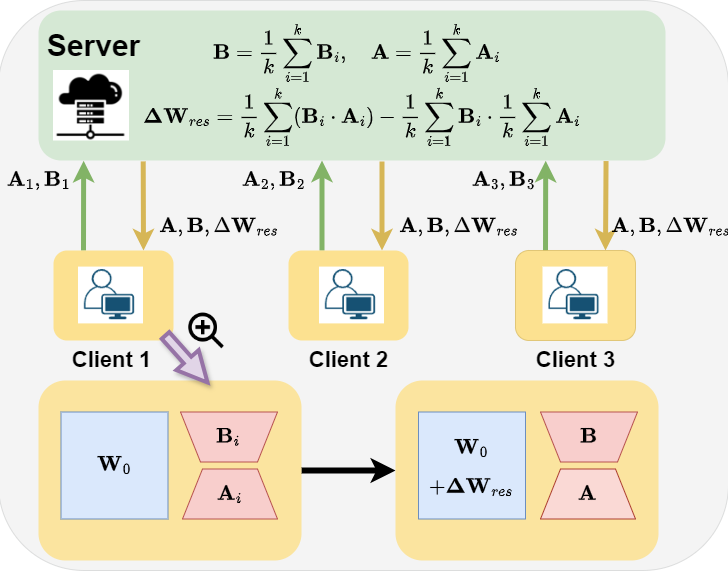}
        \caption{FedEx-LoRA}
        \label{fig:rounds-agg-main}
    \end{subfigure}
    % \caption{Difference in aggregtaion  process in FedIT vs ours (FedEx-LoRA), (a) FedIT send individual client low-rank adapters $A_i$ and $B_i$ to the server which are simply averaged and sent back by the server to the clients. This leads to inexact, noisy aggregation. (b) Our method, FedEx-LoRA, sends individual client low-rank adapters $A_i$ and $B_i$ to the server. In contrast to FedIT, the client send back an additional error residual, $\Delta W_{res}$, to the clients, which is added to the frozen pre-trained weight matrix $W$, This leads to exact aggregtaion.}
    % \caption{Comparison of aggregation processes: (a) \textbf{FedIT} averages the individual client low-rank adapters $\mathbf{A}_i$ and $\mathbf{B}_i$ to send back to the clients, resulting in inexact, noisy updates. (b) \textbf{FedEx-LoRA} sends individual adapters $\mathbf{A}_i$ and $\mathbf{B}_i$ along with an error residual $\mathbf{\Delta W}_{res}$ to the clients, which is added to the frozen pre-trained weight matrix $W$, achieving exact aggregation.}
    \caption{Comparison of federated LoRA methods: (a) \textbf{FedIT} averages the individual client low-rank adapters $\mathbf{A}_i$ and $\mathbf{B}_i$, resulting in inexact updates. (b) \textbf{FedEx-LoRA} sends the error residual $\mathbf{\Delta W}_{res}$ along with the individual adapters $\mathbf{A}_i$ and $\mathbf{B}_i$, which is added to the pretrained weight matrix $\mathbf{W}_0$, ensuring exact aggregation. Clients transmit low-rank adapters $\mathbf{A}_i$ and $\mathbf{B}_i$ in both methods.}

    \label{fig:main}
\end{figure}

The aggregation process must be carefully designed for both accuracy and simplicity. We introduce FedEx-LoRA, a method that improves federated aggregation for LoRA by incorporating an error residual term, $\mathbf{\Delta W}_{res}$, into the pretrained weight matrix to address inexact aggregation, as shown in Figure \ref{fig:main}. This adjustment preserves the low-rank efficiency of LoRA without adding computational overhead. Since the average update is inherently higher rank and cannot fit into the low-rank adapters, it is absorbed into the pretrained weight matrix, which is already high rank. This error term requires no training and is added at each aggregation step, ensuring no additional training costs.

Our key contributions are summarized as follows:
 \begin{itemize}[left=0pt]
     \item We identify a critical discrepancy in traditional federated averaging of LoRA adapters and address it by explicitly assigning the error residual to the pretrained weight matrix, ensuring ideal updates.
     \item The error residual term is incorporated at each aggregation step, maintaining LoRA’s efficiency without any additional training. We propose a communication protocol that minimizes both communication and computational overhead. 
     \item We demonstrate the effectiveness of our approach through extensive experiments on models ranging from RoBERTa-base (125M) to Gemma-2 (9B) across arithmetic reasoning, commonsense reasoning, natural language understanding, and generation tasks. Our method consistently outperforms state-of-the-art federated fine-tuning techniques, showing clear performance gains.
     \item We provide a detailed analysis of the deviations introduced by federated averaging compared to ideal updates, and identify notable patterns. We further show that while multiple assignment strategies exist for exact aggregation, our specific assignment approach is most effective.
 \end{itemize} 

\section{Related work}

\textbf{Parameter-efficient Fine-tuning.}
PEFT methods aim to adapt foundation models while minimizing the number of trainable parameters. Input-based techniques like prefix tuning \citep{prefix_tuning} prepend trainable prompts, and prompt tuning \citep{prompt_tuning} optimizes soft prompts in the embedding space - both effective for task-specific adaptations. Architectural approaches, such as adapter layers \citep{adaptor_layer}, add trainable components between transformer blocks \citep{vaswanietal}, facilitating multi-task learning. LoRA \citep{lora} reduces memory overhead by representing weight updates with low-rank matrices, while AdaLoRA \citep{adalora} improves efficiency by dynamically adjusting the parameter budget. Optimization techniques, like QLoRA \citep{qlora}, enable fine-tuning on consumer hardware via quantization, and LongLoRA \citep{longlora} targets long-context tasks. Recent advancements include combining multiple PEFT methods \citep{merge_peft} and scaling these techniques for very large models \citep{scaling_llm}, advancing the state of efficient fine-tuning.

\textbf{Federated Fine-Tuning of Foundation Models.}
Federated learning \citep{konečný2017federatedlearningstrategiesimproving} is a decentralized approach that allows multiple clients to collaboratively train a shared model without sharing their private data. Instead, clients perform local training on their own datasets, and only the resulting model updates are securely aggregated to update the global model \citep{kairouz2021advances}. This iterative process of local training and global aggregation continues until the model converges.  %Key challenges in federated learning include dealing with non-IID data, communication efficiency, model personalization, straggler clients, security and privacy. Recent advancements have focused on improving communication efficiency, tackling data heterogeneity, and enhancing privacy guarantees through techniques like secure aggregation and differential privacy.
FedBERT \citep{tian2022fedbert} introduced federated pre-training for BERT, while recent efforts have focused on federated fine-tuning of foundation models \citep{zhang2022federated, kuang2024federatedscope, babakniya2023slora}. The current state-of-the-art, FedIT \citep{FedIT}), fine-tunes LLMs by averaging LoRA parameters across clients using vanilla Federated Averaging (FedAvg, \citet{mcmahan2017communication}). However, averaging low-rank adapters independently introduces noise and results in inexact global updates. \textbf{F}ederated \textbf{F}reeze \textbf{A} LoRA (FFA-LoRA) \citep{sun2024improving} mitigates this by keeping one set of adapters trainable, improving aggregation stability but limiting the training flexibility of other adapters. This method is particularly advantageous in privacy-sensitive settings \citep{huang2022large, zhang2021survey}. Another challenge arises from heterogeneous rank settings, where clients adjust LoRA ranks based on their capacities \citep{zhao2018federated, li2019convergence}. Some methods address this by self-pruning local LoRA modules and employing sparsity-weighted aggregation \citep{hetero_lora}, though this introduces substantial computational overhead.
\section{Preliminaries and Motivation} \label{sec:motivation}
\textbf{Fine-tuning with LoRA.}
LoRA \citep{lora} leverages low-rank matrix factorization to efficiently represent the updates of pre-trained model weights. Specifically, the fine-tuned weights, $\mathbf{W}'$, are expressed as a sum of the original weights $\mathbf{W_0}$ and a low-rank update $\mathbf{\Delta W}$:
\begin{align} 
    \mathbf{W}' = \mathbf{W_0} + \mathbf{\Delta W} = \mathbf{W_0} + \mathbf{B}\mathbf{A}
    \label{eq:lora}
\end{align}
where $\mathbf{W_0}, \mathbf{W}' \in \mathbb{R}^{m \times n}$ are the pretrained and fine-tuned weight matrices, respectively, and $\mathbf{A} \in \mathbb{R}^{r \times n}$, $\mathbf{B} \in \mathbb{R}^{m \times r}$ represent the low-rank decomposition of $\mathbf{\Delta W}$. Here, the rank $r$ is significantly smaller than both $m$ and $n$, leading to a substantial reduction in the number of trainable parameters for $\mathbf{\Delta W}$. Instead of directly updating $\mathbf{W}_0$ during fine-tuning, LoRA optimizes the smaller matrices $\mathbf{A}$ and $\mathbf{B}$, resulting in considerable savings in memory usage. For instance, in GPT-2, LoRA reduces the number of trainable parameters from $124.44$ M to just $0.41$ M when using a rank of $r=4$, with no observed degradation in performance \citep{lora}.

\textbf{Global Updates due to Vanilla Federated Averaging are Inexact.}
The widely adopted federated learning algorithm, FedAvg \citep{mcmahan2017communication}, updates the global model by performing a weighted average of local client updates in each communication round for $k$ clients:
\begin{align}
    \mathbf{W}^{global} = \mathbf{W_0} + \frac{1}{k}\sum_{i=1}^k \mathbf{\Delta W}_i = \mathbf{W_0} + \mathbf{\Delta W}
    \label{eq:fedavg}
\end{align}
where $\mathbf{W_0}$ and $\mathbf{W}^{global}$ represent the global model parameters before and after aggregation, respectively. $\mathbf{\Delta W}_i$ denotes the local update from the $i$-th client.
FedIT \citep{FedIT} extends FedAvg by incorporating LoRA for federated fine-tuning, where clients fine-tune LoRA modules of a fixed rank. 
%Clients first download a pre-trained language model from the server, locally fine-tune the LoRA modules, and then send the updated LoRA parameters back to the server. 
The global LoRA matrices $\mathbf{A}$ and $\mathbf{B}$ are updated via weighted averaging over the client-specific LoRA parameters $\mathbf{A}_k$ and $\mathbf{B}_k$:
\begin{align}
    \mathbf{A} = \frac{1}{k}\sum_{i=1}^{k}\mathbf{A}_i, \quad \mathbf{B} = \frac{1}{k}\sum_{i=1}^{k} \mathbf{B}_i
    \label{eq:fedit}
\end{align}
Although FedIT follows a similar aggregation process as FedAvg, only LoRA modules are updated and communicated. However, this independent averaging of $\mathbf{A}_i$ and $\mathbf{B}_i$ introduces deviation from the exact centralized LoRA updates, as the actual model updates depend on the product $\mathbf{B}_i \mathbf{A}_i$, not the individual components $\mathbf{B}$ and $\mathbf{A}$.
\begin{align}
    \underbrace{\Tilde{\mathbf{W}}^{global} =  \mathbf{W}_0 + \frac{1}{k} \sum_{i=1}^k \mathbf{B}_i \times \frac{1}{k} \sum_{i=1}^k \mathbf{A}_i}_{\text{Parameters after aggregation with LoRA + FedAvg (FedIT)}} \neq \underbrace{\mathbf{W}_0 + \frac{1}{k} \sum_{i=1}^k (\mathbf{B}_i \mathbf{A}_i) = \mathbf{W}^{global}}_{\text{Ideal parameters following model-averaging}}
    \label{eq:fedavg-inexact}
\end{align}
\textbf{There is No Free Lunch.} A naive approach would be to directly average the client updates as $\frac{1}{k}\sum_{i=1}^{k} (\mathbf{B}_i \mathbf{A}_i)$ and use the result for the global update before resuming training. However, this undermines the purpose of LoRA, as it forces subsequent training on the full-rank matrix $\mathbf{W}^{global} \in \mathbb{R}^{m \times n}$ rather than its intended low-rank adapters $\mathbf{A} \in \mathbb{R}^{r \times n}$ and $\mathbf{B} \in \mathbb{R}^{m \times r}$.

An alternative is to decompose the averaged update $\frac{1}{k}\sum_{i=1}^{k} (\mathbf{B}_i \mathbf{A}_i)$ into a low-rank matrix of rank $(k \cdot r)$. However, this leads to an exponential growth in the rank with each aggregation round, as the rank increases by a factor of $k$ in every iteration, making this approach computationally intractable.

\textbf{FFA-LoRA.}
FFA-LoRA addresses the problem of inexact aggregation, particularly in privacy-preserving settings. Motivated from previous works \citep{zhang2023lorafamemoryefficientlowrankadaptation, tian2024hydraloraasymmetricloraarchitecture}, it asymmetrically freezes the $\mathbf{A}$ adapters while keeping only the $\mathbf{B}$ adapters trainable. This approach mitigates the issues of non-ideal aggregation by avoiding independent updates of $\mathbf{A}$ and $\mathbf{B}$. However, the drawback is that the $\mathbf{A}$ matrix remains static, which limits expressiveness. While this method excels in privacy-sensitive scenarios where noise is amplified, it underperforms in non-private settings, even when the number of trainable parameters is equivalent.
\section{Method: FedEx-LoRA}\label{sec:method}

\subsection{Noise-Free Exact Aggregation}
% To solve this issue of noisy inexact aggregation caused by averaging A and B matrices separately, we propose a method.
% We then average product of BA
% however, as mentioned we cannot keep this entire matrix (or decomposition) trainable
% thus, we append the error between product of average of A and B versus average of product of A and B. this is a high rank matrix  
% Our solution includes appending this high-rank update matrix in the global frozen weight matrix. this high rank matrix is not trainable, we simply add the error residual term in each aggregation
To tackle the problem of inexact aggregation arising from the independent averaging of the $\mathbf{A}$ and $\mathbf{B}$ matrices across clients, we introduce a novel method called FedEx-LoRA. Instead of separately averaging the low-rank adapter matrices $\mathbf{A}$ and $\mathbf{B}$, we compute the average of their product $\mathbf{BA}$ across all clients. However, as previously noted in Section \ref{sec:motivation}, we cannot keep this high-rank matrix or its lower-rank decomposition (with rank $(k\cdot r)$) trainable. Consequently, we append a high-rank error term that captures the discrepancy between the average of the products and the product of the averages. This error residual is incorporated into the global frozen weight matrix, ensuring its non-trainability. The update at the $j^{th}$ aggregation round can be expressed as follows:
\begin{gather}
    \mathbf{B}^{j+1}_i \leftarrow \frac{1}{k} \sum_{i=1}^k \mathbf{B}^{j}_i, \quad \mathbf{A}^{j+1}_i \leftarrow \frac{1}{k} \sum_{i=1}^k \mathbf{A}^{j}_i \label{eq:ass-ab} \\
    \mathbf{W_0}^{j+1} \leftarrow \mathbf{W_0}^{j} + \underbrace{\frac{1}{k} \sum_{i=1}^k (\mathbf{B}^{j}_i \mathbf{A}^{j}_i) - \frac{1}{k} \sum_{i=1}^k \mathbf{B}^{j}_i \times \frac{1}{k} \sum_{i=1}^k \mathbf{A}^{j}_i}_{\text{Residual}} 
\end{gather}
We now demonstrate that our formulation results in exact aggregation for every client:
\begin{gather}
     \mathbf{W}_{global}^{j+1} = \mathbf{W_0}^{j} + \mathbf{B}_i^{j} \mathbf{A}_i^{j} \label{eq:prove-1}\\
     \mathbf{W}_{global}^{j+1} = \mathbf{W_0}^{j} + \frac{1}{k} \sum_{i=1}^k (\mathbf{B}_i^{j} \mathbf{A}_i^{j}) - \frac{1}{k} \sum_{i=1}^k \mathbf{B}_i^{j} \times \frac{1}{k} \sum_{i=1}^k \mathbf{A}_i^{j} + \frac{1}{k} \sum_{i=1}^k \mathbf{B}_i^{j} \times \frac{1}{k} \sum_{i=1}^k \mathbf{A}_i^{j} \label{eq:prove-2} \\
     \mathbf{W}_{global}^{j+1} = \underbrace{\mathbf{W_0}^{j} + \frac{1}{k} \sum_{i=1}^k (\mathbf{B}_i^{j} \mathbf{A}_i^{j})}_{\text{Ideal aggregation}} \label{eq:prove-2}
\end{gather}

\subsection{FedEx-LoRA: Overall Pipeline}
Initially, the server distributes the global pretrained model to all \( k \) clients and initializes the low-rank adapters \( \mathbf{A} \) and \( \mathbf{B} \) according to standard LoRA settings: \( \mathbf{B} \) is initialized to zero, while \( \mathbf{A} \) is initialized using a random Gaussian distribution.
\begin{align}
        \mathbf{B}_i^{0} \leftarrow \mathbf{B}_{init}, \quad \mathbf{A}_i^{0} \leftarrow \mathbf{A}_{init}, \quad \mathbf{W}_0^{0} \leftarrow \mathbf{W}_{pretrained}
\end{align}
Each client then independently trains their low-rank adapters \( \mathbf{A} \) and \( \mathbf{B} \) using their local data for a specified number of epochs (referred to as ``local epochs''). Upon completion of training, the clients send their updated low-rank adapters back to the server for aggregation. The server aggregates these low-rank adapters and incorporates the residual term into the global model:
\begin{gather}
 \mathbf{B}_{global}^{j} = \frac{1}{k} \sum_{i=1}^k \mathbf{B}_i^{j}, \quad\mathbf{A}_{global}^{j}= \frac{1}{k} \sum_{i=1}^k \mathbf{A}_i^{j} \\
    \mathbf{ \Delta W}_{res}^{j} = \frac{1}{k} \sum_{i=1}^k (\mathbf{B}_i^{j} \mathbf{A}_i^{j}) - \frac{1}{k} \sum_{i=1}^k \mathbf{B}_i^{j} \times \frac{1}{k} \sum_{i=1}^k \mathbf{A}_i^{j} 
\end{gather}

% \begin{gather}

% \end{gather}
The server then sends the aggregated matrices back to each client. After receiving these updates, the clients proceed to update their low-rank adapters \( \mathbf{A} \) and \( \mathbf{B} \), as well as the weight matrix:
\begin{gather}
        \mathbf{B}_i^{j+1} \leftarrow \mathbf{B}^{j}_{global}, \quad \mathbf{A}_i^{j+1} \leftarrow \mathbf{A}^{j}_{global} \\
        \mathbf{W}_0^{j+1} \leftarrow \mathbf{W}_0^{j} + \mathbf{\Delta W}_{res}^{j}
\end{gather}
Following this, clients independently resume fine-tuning for a set number of local epochs. This process repeats across multiple aggregation rounds (also referred to as communication rounds).

\textbf{Multiple Assignment Strategies can Lead to Exact Aggregation.}
Several methods can be used for achieving exact aggregation, with our choice of assignments for \( \mathbf{A}_i \) and \( \mathbf{B}_i \) being particularly pivotal. Each such assignment strategy allows us to adjust the corresponding error offset within the frozen weight matrix, facilitating precise aggregation. In Section \ref{anal-inits}, we investigate various methods and empirically show that our proposed assignments for \( \mathbf{A}_i \) and \( \mathbf{B}_i \)  deliver the best performance.

\textbf{Communication Protocol.}
At first glance, it may seem necessary for the server to transmit the high-rank update matrix $\mathbf{\Delta W}_{res}$ to the clients, which could introduce substantial communication overhead. However, the rank of this update matrix is capped at $(k \cdot r)$. 
Consequently, $\mathbf{\Delta W}_{res}$ can be decomposed into two low-rank matrices using methods such as Gram-Schmidt orthogonalization. This decomposition expresses the matrix as a product of the basis of its column (or row) space and the corresponding linear coefficients. 
The \textit{computational} overhead incurred by this operation at each aggregation step is negligible compared to the numerous matrix multiplications involved in training.
Importantly, clients are only required to transmit their low-rank adapters $\mathbf{A}_i$ and $\mathbf{B}_i$, avoiding the need to send any high-rank update matrices. 
In practice, the \textit{communication} overhead is minimal compared to FedIT, and overall, the \textit{communication} cost remains significantly lower than that of full federated fine-tuning. Detailed communication overhead analysis is provided in Section \ref{sec:analysis}.

% \textbf{Best Inexact Approximation.}
% The communication cost for exact aggregation scales linearly with number of clients. This cost can be unrealistic to bear for hyperclient settings. We propose a method to mitigate this issue, by relaxing the exact aggregation condition .We take the truncated SVD of the resdiual matrix, to reconstruct its low rank approximation. This is the most optimum low-rank approximation of the high rank update matrix, by the Eckart-Young theorem. The best low rank approximation $\Delta W_{rec}^{r'}$ of rank say $r'$ is given  y:
% \begin{gather}
%     U,S,V^T \leftarrow \textbf{SVD}(\Delta W_{res}) \label{W_svd}\\
%     \Delta W_{rec}^{r'}  \leftarrow U[1:r'] S[1:r',1:r']  V^T[1:r'] \label{S_init}   
% \end{gather}

% This method, while inexact, is the best approximation of the exact aggregation. Moreover, this allows the server to dictate the communication cost, an option not present in the previous methods \textbf{FedIT} \citep{FedIT}, and \textbf{FFA-LoRA} \citep{sun2024improving}.
\textbf{Best Inexact Approximation.}
For exact aggregation, the communication cost scales linearly with the number of clients, becoming prohibitive in hyperclient settings. To address this, we propose relaxing the exact aggregation condition through truncated SVD of the residual matrix. This reconstruction yields a low-rank approximation which, by the Eckart-Young theorem \citep{eckart1936theorem1}, is provably optimal for the high-rank update matrix. Specifically, for a target rank $r'$, the best low-rank approximation $\Delta W_{rec}^{r'}$ is computed as:
\begin{gather}
    U,S,V^T \leftarrow \textbf{SVD}(\Delta W_{res}) \label{W_svd}\\
    \Delta W_{rec}^{r'}  \leftarrow U[1:r'] S[1:r',1:r']  V^T[1:r'] \label{S_init}   
\end{gather}
While this method introduces approximation error, it provides the theoretically optimal approximation to exact aggregation. A key advantage is that the server can control communication costs, a capability absent in previous methods - FedIT \citep{FedIT} and FFA-LoRA \citep{sun2024improving}.

% \begin{itemize}
%     \item After each round, all clients send their low-rank adaptors, $\mathbf{A}_i$ and $\mathbf{B}_i$, to the server.
%     \item The server performs aggregation and computes $\Delta \mathbf{W}_{res}$, $\mathbf{A}^{global}$, and $\mathbf{B}^{global}$.
%     \item The server then decomposes the residual update matrix into two matrices $\mathbf{W}_1$ and $\mathbf{W}_2$ of sizes $(n, m \cdot r)$ and $(m \cdot r, n)$, respectively, such that $\Delta \mathbf{W}_{res} = \mathbf{W}_1 \cdot \mathbf{W}_2$.
%     \item Subsequently, the server sends these three matrices to each client.
%     \item Clients reconstruct $\Delta \mathbf{W}_{res}$ using $\mathbf{W}_1$ and $\mathbf{W}_2$ and add it to the frozen $\mathbf{W}$ matrix. The two low-rank adaptors are then initialized using $\mathbf{A}^{global}$ and $\mathbf{B}^{global}$ for further training.
% \end{itemize}

\section{Experiments}\label{sec:experiments}
%\vspace{-1mm}
\textbf{Models and Datasets.}
We evaluate our method on four NLP benchmarks using models ranging from RoBERTa-base with 125M parameters to Gemma-2 with 9B parameters, covering both masked and autoregressive architectures. Our experiments include fine-tuning Mistral-7B \citep{mistral7b}, Gemma-2 9B \citep{gemma2}, Llama-3.2 3B \citep{llama3}, RoBERTa-base, RoBERTa-large \citep{liu2019robertarobustlyoptimizedbert}, and GPT-2 \citep{radford2019language} using FedEx-LoRA.  This comprehensive setup allows us to assess the effectiveness of our approach across different tasks and model architectures.

For arithmetic reasoning, we fine-tune the decoder-only models Mistral-7B and Gemma-2 9B using 10K samples from the MetaMathQA dataset \citep{metamathqa}. These models are evaluated on two standard arithmetic reasoning benchmarks, GSM8K \citep{gsm8k} and MATH \citep{math}. In the commonsense reasoning category, we use Llama-3.2 3B, which is trained on \textsc{CommonSense170K}—a compilation of eight commonsense reasoning datasets \citep{cr-dataset}.
We evaluate the RoBERTa models on natural language understanding tasks with the GLUE benchmark \citep{wang2019gluemultitaskbenchmarkanalysis} and assess GPT-2 on natural language generation tasks through the E2E NLG Challenge \citep{novikova2017e2e}. We implement all algorithms using PyTorch \citep{paszke2019pytorch}, based on the widely-used HuggingFace Transformers codebase \citep{wolf2020transformers}. We run all experiments on a single NVIDIA A100/A6000 GPU, and present the results as average of $3$ different random runs. Base models are loaded in \texttt{\textbf{torch.bfloat16}} to save memory. Dataset details are presented in Appendix \ref{app:datasets}.

\textbf{Implementation Details.} 
The residual and product matrices are scaled by the factor $\alpha/r$, where $\alpha$ is a constant in $r$, consistent with the approach in LoRA \citep{lora}. We run our experiments in a three-client cross-silo federated setting, based on the settings described in FFA-LoRA \citep{sun2024improving}. For data distribution among clients, we use the common method to sample data at random for each client, as implemented in standard works \citep{FedIT, he2020fedml, lai2022fedscale}.

\textbf{Baselines.}
We primarily compare FedEx-LoRA with other federated fine-tuning versions of LoRA, but include centralized LoRA as a \textit{performance benchmark} or \textit{skyline}. We also include other baselines, where possible. \textbf{Full Fine-Tuning (FT)} refers to fine-tuning the entire pretrained model. \textbf{LoRA} \citep{lora} represents the traditional centralized LoRA approach. \textbf{FedIT} \citep{FedIT}, the current state-of-the-art federated fine-tuning method, applies vanilla federated averaging (FedAvg) to LoRA \citep{mcmahan2017communication}. \textbf{FFA-LoRA} \citep{sun2024improving} freezes the $\mathbf{A}$ matrices and trains only the $\mathbf{B}$ matrices, allowing for exact aggregation in a federated setting but at the cost of losing the benefits of training $\mathbf{A}$.

\subsection{Instruction Tuning}

\textbf{Implementation Details.} 
For \textbf{arithmetic reasoning}, we fine-tune Mistral-7B \citep{mistral7b} and Gemma-2 9B \citep{gemma2} on 10K samples from the MetaMathQA dataset \citep{metamathqa} and evaluate them on the GSM8K \citep{gsm8k} and MATH \citep{math} benchmarks. For \textbf{commonsense reasoning}, we use Llama-3.2 3B, training it on \textsc{CommonSense170K}—a dataset combining eight commonsense reasoning datasets \citep{cr-dataset}—and evaluate its performance on each of those datasets. In all instruction tuning tasks, we apply LoRA modules to the key, value, query, attention output, and all fully connected weight matrices. We fine-tune over a single local epoch within one aggregation round, using a rank of $r=32$.

\textbf{Main Results.}  
Tables \ref{tab:commonsense} and \ref{tab:math} present the results for commonsense and arithmetic reasoning. Our method consistently surpasses state-of-the-art federated fine-tuning techniques across both arithmetic benchmarks and all eight commonsense reasoning tasks for every evaluated model. For example, on average accuracy for commonsense reasoning, FedEX-LoRA outperforms FFA-LoRA by $8.63\%$ and FedIT by $2.42\%$ respectively.

\begin{table}[ht]
\centering
\setlength{\tabcolsep}{4pt}
\small
\begin{tabular}{l|ccccccccc}
\hline
\toprule
\multirow{2}*{\bf Method} & \multicolumn{8}{c}{\textbf{Accuracy ($\uparrow$)}} & \\
& \textbf{BoolQ} & \textbf{PIQA} & \textbf{SIQA} & \textbf{HellaS.} & \textbf{WinoG.} & \textbf{ARC-e} & \textbf{ARC-c} & \textbf{OBQA} & \textbf{Avg.} \\
\midrule
\rowcolor{gray!20} $\text{Centralized LoRA}_{\text{r}=32}$ & $73.45$ & $89.65$ & $82.23$ & $94.41$ & $87.97$ & $93.88$ & $82.76$ & $86.60$ & $86.37$ \\
\rowcolor{cyan!10}$\text{FedIT}_{\text{r}=32}$ & $70.73$ & $87.59$ & $79.17$ & $91.06$ & $83.42$ & $92.71$ & $81.31$ & $82.68$ & $83.57$ \\
\rowcolor{cyan!10}$\text{FFA-LoRA}_{\text{r}=32}$ & $65.78$ & $84.22$ & $72.41$ & $82.27$ & $72.53$ & $90.36$ & $76.28$ & $75.00$ & $77.35$ \\
\rowcolor{cyan!10}$\text{FedEx-LoRA}_{\text{r}=32}$ & $\mathbf{73.21}$ & $\mathbf{89.01}$ & $\mathbf{81.98}$ & $\mathbf{94.29}$ & $\mathbf{87.29}$ & $\mathbf{93.68}$ & $\mathbf{82.33}$ & $\mathbf{86.20}$ & $\mathbf{85.99}$ \\
\bottomrule
\end{tabular}
\caption{Results for Llama-3.2 3B on eight commonsense reasoning datasets, comparing various federated LoRA methods at rank $r=32$. \textbf{Centralized LoRA (in grey) sets the benchmark skyline} for its federated versions. Best results among federated methods (in blue) are highlighted in \textbf{bold} for each setting.}
\label{tab:commonsense}
\end{table}

\begin{table}[ht]
\centering
\small
\begin{tabular}{ll|cc}
\toprule
\multirow{2}{*}{\textbf{Model}} & \multirow{2}{*}{\textbf{Method}} & \multicolumn{2}{c}{\textbf{Accuracy ($\uparrow$)}} \\
\cmidrule{3-4}
& & \textbf{GSM8K} & \textbf{MATH} \\
\midrule
\multirow{4}{*}{Mistral-7B} 
    & \cellcolor{gray!20}$\text{Centralized LoRA}_{\text{r}=32}$ & \cellcolor{gray!20}$62.77$ & \cellcolor{gray!20}$16.24$ \\
    & \cellcolor{cyan!10}$\text{FedIT}_{\text{r}=32}$ & \cellcolor{cyan!10}$56.94$ & \cellcolor{cyan!10}$14.96$ \\
    & \cellcolor{cyan!10}$\text{FFA-LoRA}_{\text{r}=32}$ & \cellcolor{cyan!10}$56.41$ & \cellcolor{cyan!10}$14.88$ \\
    & \cellcolor{cyan!10}$\text{FedEx-LoRA}_{\text{r}=32}$ & \cellcolor{cyan!10}$\mathbf{62.62}$ & \cellcolor{cyan!10}$\mathbf{16.54}$ \\
\midrule
\multirow{4}{*}{Gemma-2 9B} 
    & \cellcolor{gray!20}$\text{Centralized LoRA}_{\text{r}=32}$ & \cellcolor{gray!20}$76.34$ & \cellcolor{gray!20}$39.32$ \\
    & \cellcolor{cyan!10}$\text{FedIT}_{\text{r}=32}$ & \cellcolor{cyan!10}$74.57$ & \cellcolor{cyan!10}$37.16$ \\
    & \cellcolor{cyan!10}$\text{FFA-LoRA}_{\text{r}=32}$ & \cellcolor{cyan!10}$75.04$ & \cellcolor{cyan!10}$35.18$ \\
    & \cellcolor{cyan!10}$\text{FedEx-LoRA}_{\text{r}=32}$ & \cellcolor{cyan!10}$\mathbf{76.19}$ & \cellcolor{cyan!10}$\mathbf{39.00}$ \\
\bottomrule
\end{tabular}
\caption{Arithmetic reasoning performance on GSM8K and MATH for Mistral-7B and Gemma-2 9B, comparing various federated LoRA methods at rank $r= 32$. \textbf{Centralized LoRA (in grey) sets the benchmark skyline} for its federated versions. Best results among federated methods (in blue) are highlighted in \textbf{bold} for each setting.}
\label{tab:math}
\end{table}

\subsection{Natural Language Understanding}

\textbf{Implementation Details.}
RoBERTa \citep{liu2019robertarobustlyoptimizedbert} is a widely used pretrained model known for its competitive performance among its size. We use the pretrained RoBERTa-base (125M parameters) and RoBERTa-large (355M parameters) from the HuggingFace Transformers library \citep{wolf2020transformers} and evaluate them on several datasets from the GLUE benchmark: CoLA, RTE, MRPC, SST-2, QNLI, and STS-B. We apply LoRA modules only to the self-attention layers, following the setup from the original LoRA paper \citep{lora}. Models are fine-tuned at ranks $r=\{4,1\}$ over local epochs of $3$ and $10$. For RoBERTa-base, we run $50$ aggregation rounds for $3$ local epochs and $15$ rounds for $10$ local epochs. For RoBERTa-large, we perform $15$ aggregation rounds for $3$ local epochs and $5$ rounds for $10$ local epochs. Detailed experimental settings are provided in Appendix \ref{app:hyperparams}. 

\textbf{Main Results.}
We present results for RoBERTa-base and RoBERTa-large in Table \ref{tab:gluedatasets}, evaluated at ranks $r = \{4,1\}$. Our method consistently outperforms state-of-the-art federated fine-tuning approaches across all datasets and settings. Notably, our method occasionally achieves performance on par with centralized LoRA. Additional results in Appendix \ref{app:nlu} (Table \ref{tab:app-gluedatasets}) further demonstrate the robustness and superiority of our method over other federated LoRA variants across multiple settings.

\begin{table}[ht]
    \begin{center}
        
        \begin{minipage}{\textwidth}
            \centering
            \small
            
            \begin{tabular}{l|ccccccc}
                \toprule
                \multirow{2}{*}{\bf Method} & {\bf CoLA} & {\bf RTE} & {\bf MRPC} & {\bf SST-2} & {\bf QNLI} & {\bf STS-B} & {\bf All} \\
                ~ & {Mcc $\uparrow$} & {Acc $\uparrow$} & {Acc $\uparrow$} & {Acc $\uparrow$} & {Acc $\uparrow$} & {Corr $\uparrow$} & {Avg $\uparrow$} \\
                \midrule
                \rowcolor{gray!20} $\text{Centralized LoRA}_{r=4}$ & $64.31$ & $75.45$ & $87.99$ & $94.61$ & $92.75$ & $90.73$ & $84.31$ \\
                \rowcolor{cyan!10}$\text{FedIT}_{r=4}$ & $60.82$ & $73.64$ & $88.48$ & $94.61$ & $92.07$ & $90.91$ & $83.42$ \\
                \rowcolor{cyan!10}$\text{FFA-LoRA}_{r=4}$ & $59.34$ & $70.04$ & $87.50$ & $94.27$ & $91.37$ & $90.26$ & $82.13$ \\
                \rowcolor{cyan!10} $\text{FedEx-LoRA}_{r=4}$ & $\mathbf{62.82}$ & $\mathbf{75.09}$ & $\mathbf{89.95}$ & $\mathbf{94.84}$ & $\mathbf{92.66}$ & $\mathbf{90.95}$ & $\mathbf{84.39}$ \\
                \midrule
                \rowcolor{gray!20} $\text{Centralized LoRA}_{r=1}$ & $62.13$ & $74.67$ & $87.75$ & $94.61$ & $92.31$ & $90.83$ & $83.72$ \\
                \rowcolor{cyan!10}$\text{FedIT}_{r=1}$ & $61.33$ & $71.48$ & $87.99$ & $94.52$ & $92.01$ & $90.81$ & $83.02$ \\
                \rowcolor{cyan!10}$\text{FFA-LoRA}_{r=1}$ & $57.52$ & $71.20$ & $87.48$ & $94.03$ & $91.78$ & $90.34$ & $82.06$ \\
                \rowcolor{cyan!10}$\text{FedEx-LoRA}_{r=1}$ & $\mathbf{62.07}$ & $\mathbf{73.65}$ & $\mathbf{88.73}$ & $\mathbf{94.84}$ & $\mathbf{92.21}$ & $\mathbf{90.87}$ & $\mathbf{83.73}$ \\
                \bottomrule
            \end{tabular}
            \caption*{(a) Results with RoBERTa-base on the GLUE benchmark datasets}
        \end{minipage}

        \vspace{0.5cm} % Space between subtables

        \begin{minipage}{\textwidth}
            \centering
            \small
            \begin{tabular}{l|ccccccc}
                \toprule
                \multirow{2}{*}{\bf Method} & {\bf CoLA} & {\bf RTE} & {\bf MRPC} & {\bf SST-2} & {\bf QNLI} & {\bf STS-B} & {\bf All} \\
                ~ & {Mcc $\uparrow$} & {Acc $\uparrow$} & {F1 $\uparrow$} & {Acc $\uparrow$} & {Acc $\uparrow$} & {Corr $\uparrow$} & {Avg $\uparrow$} \\
                \midrule
                \rowcolor{gray!20} $\text{Centralized LoRA}_{r=4}$ & $66.03$ & $82.67$ & $88.84$ & $96.21$ & $94.58$ & $91.92$ & $86.71$ \\
                \rowcolor{cyan!10}$\text{FedIT}_{r=4}$ & $64.48$ & $78.43$ & $88.48$ & $95.87$ & $94.41$ & $91.29$ & $85.49$ \\
                \rowcolor{cyan!10}$\text{FFA-LoRA}_{r=4}$ & $62.05$ & $75.39$ & $86.52$ & $95.27$ & $94.35$ & $90.23$ & $83.97$ \\
                \rowcolor{cyan!10}$\text{FedEx-LoRA}_{r=4}$ & $\mathbf{65.29}$ & $\mathbf{80.31}$ & $\mathbf{89.95}$ & $\mathbf{96.21}$ & $\mathbf{94.71}$ & $\mathbf{91.85}$ & $\mathbf{86.39}$ \\
                \midrule
                \rowcolor{gray!20} $\text{Centralized LoRA}_{r=1}$ & $65.21$ & $83.39$ & $92.44$ & $96.10$ & $94.42$ & $92.12$ & $87.28$ \\
                \rowcolor{cyan!10}$\text{FedIT}_{r=1}$ & $62.82$ & $78.11$ & $91.29$ & $96.10$ & $94.35$ & $91.62$ & $85.72$ \\
                \rowcolor{cyan!10}$\text{FFA-LoRA}_{r=1}$ & $60.58$ & $74.67$ & $89.47$ & $95.58$ & $94.01$ & $91.34$ & $84.28$ \\
                \rowcolor{cyan!10}$\text{FedEx-LoRA}_{r=1}$ & $\mathbf{64.35}$ & $\mathbf{80.01}$ & $\mathbf{91.76}$ & $\mathbf{96.22}$ & $\mathbf{94.71}$ & $\mathbf{91.91}$ & $\mathbf{86.49}$ \\
                \bottomrule
            \end{tabular}
            \caption*{(b) Results with RoBERTa-large on the GLUE benchmark datasets}
        \end{minipage}

        \caption{Results with RoBERTa-base and Roberta-large on the GLUE benchmark datasets, comparing various federated LoRA methods at ranks $r= \{4,1\}$. \textbf{Centralized LoRA (in grey) sets the benchmark skyline} for its federated versions. Best results among federated methods (in blue) are highlighted in \textbf{bold} for each setting. There are $3$ local epochs before every aggregation round. We report Matthew's correlation for CoLA, Pearson correlation for STS-B, and accuracy for others. Higher is better for all metrics.}
        \label{tab:gluedatasets}
        
    \end{center}
\end{table}

\subsection{Natural Language Generation}

\textbf{Implementation Details.} 
We fine-tune GPT-2 (124M parameters) \citep{radford2019language} on the E2E NLG Challenge dataset \citep{novikova2017e2e}.  We apply LoRA modules only to the self-attention layers. The model is fine-tuned at ranks $r=\{4,1\}$ with local epochs set to $3$ and $10$, using $6$ aggregation rounds for both settings. Detailed experimental settings are provided in Appendix \ref{app:hyperparams}.

\textbf{Main Results.} 
Table \ref{tab:gpt2_ft_results} presents the performance of GPT-2 fine-tuned with ranks $r = \{4,1\}$. FedEx-LoRA consistently outperforms leading federated fine-tuning methods, across all metrics and settings. 
% Notably, as shown in Table \ref{tab:gpt2_ft_results}, our approach matches the performance of centralized full model fine-tuning as well. 
Additional evaluations, provided in Appendix \ref{app:nlg} (Table \ref{tab:gpt2_ft_results-app}), further demonstrate the reliability and strength of FedEx-LoRA across different configurations.

\begin{table}[ht]
\centering
\small
\begin{tabular}{l|ccccc}
\hline
\toprule
\multirow{2}*{\bf Method} & \multicolumn{5}{c}{\textbf{E2E NLG Challenge}} \\
       & BLEU $\uparrow$ & NIST $\uparrow$ & MET $\uparrow$ & ROUGE-L $\uparrow$ & CIDEr $\uparrow$\\
\midrule
% \rowcolor{gray!20} Full Fine-Tuning *  & $68.20$ & $8.62$ & $46.20$ & $71.00$ & $2.47$  \\
% \midrule
\rowcolor{gray!20} $\text{Centralized LoRA}_{\text{r}=4}$  & $68.91$ & $8.73$ & $46.78$ & $71.29$ & $2.47$  \\
\rowcolor{cyan!10}$\text{FedIT}_{\text{r}=4}$ & $67.60$ & $8.67$ & $46.30$ & $68.96$ & $2.41$  \\
\rowcolor{cyan!10}$\text{FFA-LoRA}_{\text{r}=4}$ & $66.79$ & $8.61$ & $45.24$ & $67.98$ & $2.39$  \\
\rowcolor{cyan!10}$\text{FedEx-LoRA}_{\text{r}=4}$ & $\mathbf{68.15}$ & $\mathbf{8.72}$ & $\mathbf{46.48}$ & $\mathbf{69.49}$ & $\mathbf{2.44}$  \\
\midrule
\rowcolor{gray!20} $\text{Centralized LoRA}_{\text{r}=1}$  & $67.41$ & $8.68$ & $46.01$ & $69.51$ & $2.41$  \\
\rowcolor{cyan!10}$\text{FedIT}_{\text{r}=1}$ & $66.01$ & $8.56$ & $45.21$ & $68.14$ & $2.28$  \\
\rowcolor{cyan!10}$\text{FFA-LoRA}_{\text{r}=4}$ & $65.87$ & $8.54$ & $45.02$ & $68.05$ & $2.27$  \\
\rowcolor{cyan!10}$\text{FedEx-LoRA}_{\text{r}=1}$ & $\mathbf{67.02}$ & $\mathbf{8.61}$ & $\mathbf{45.99}$ & $\mathbf{69.52}$ & $\mathbf{2.38}$  \\
\bottomrule
\end{tabular}
\caption{Results with GPT-2 on the E2E NLG Challenge, comparing various federated LoRA methods at ranks $r= \{4,1\}$. \textbf{Centralized LoRA (in grey) sets the benchmark skyline} for its federated versions. Best results among federated methods (in blue) are highlighted in \textbf{bold} for each setting. There are $3$ local epochs before every aggregation round. Higher is better for all metrics.}
\label{tab:gpt2_ft_results}
\end{table}

\section{Analysis}\label{sec:analysis}

To fully understand the implications of our method, we performed several in-depth analyses, each targeting a specific aspect of FedEx-LoRA's performance and efficiency.

\textbf{Assignment Strategies for $\mathbf{A}_i$ and $\mathbf{B}_i$.} \label{anal-inits}
As discussed in Section \ref{sec:method}, we can incorporate any high-rank update matrix $\mathbf{\Delta W}_{res}$ within the frozen full-rank matrix $\mathbf{W}_0$. However, assignment of the low-rank adapters $\mathbf{A}_i$ and $\mathbf{B}_i$ post-aggregation is less straightforward. 
Any selection of $\mathbf{A}_i$ and $\mathbf{B}_i$ can be offset by adjusting the residual update, by ensuring that $\mathbf{W}_0 + \mathbf{B}_i\mathbf{A}_i$ remains consistent across clients. 
We evaluate three strategies: (1) \textbf{Reinitialize $\mathbf{A}_i$ and $\mathbf{B}_i$} reinitializes $\mathbf{A}_i$ and $\mathbf{B}_i$ after aggregation and appends the full update to the frozen weights (ensuring $\mathbf{W}_0 + \mathbf{B}_i\mathbf{A}_i$ is identical). (2) $\mathbf{A}_i \leftarrow \mathbf{A}_i$ and $\mathbf{B}_i \leftarrow \mathbf{B}_i$ leaves $\mathbf{A}_i$ and $\mathbf{B}_i$ unchanged across clients, maintaining their pre-aggregation values. (3) \textbf{FedEx-LoRA} aggregates $\mathbf{A}_i$ and $\mathbf{B}_i$ using the aggregation method in FedIT (FedAvg), providing the best low-rank approximation to the aggregated update with the residual $\mathbf{\Delta W}_{res}$ stored in $\mathbf{W}_0$. We present results for RoBERTa-base on the GLUE benchmark in Table \ref{tab:init_gluedatasets}. FedEx-LoRA outperforms the other strategies, leading us to adopt $\mathbf{B}_i \leftarrow \frac{1}{k} \sum_{i=1}^k \mathbf{B}_i$ and $\mathbf{A}_i \leftarrow \frac{1}{k} \sum_{i=1}^k \mathbf{A}_i$ across all clients.

\begin{table}[ht]
\begin{center}
%\begin{small}
\begin{tabular}{l|ccccccc}
\toprule
\multirow{2}{*}{\bf Method} & {\bf CoLA} & {\bf RTE} & {\bf MRPC} & {\bf SST-2} & {\bf QNLI} & {\bf STS-B} & {\bf All} \\
~ & {Mcc $\uparrow$} & {Acc $\uparrow$} & {Acc $\uparrow$} & {Acc $\uparrow$} & {Acc $\uparrow$} & {Corr $\uparrow$} & {Avg $\uparrow$} \\
\midrule
Reinitialize $\mathbf{A}_i$ and $\mathbf{B}_i$ & $0.00$ & $61.37$ & $75.74$ & $76.26$ & $53.98$ & $53.38$ & $53.46$ \\
$\mathbf{A}_i \leftarrow \mathbf{A}_i$ and $\mathbf{B}_i \leftarrow \mathbf{B}_i$ & $55.54$ & $59.93$ & $84.80$ & $92.77$ & $88.98$ & $88.41$ & $78.41$ \\
FedEx-LoRA & $\mathbf{62.82}$ & $\mathbf{75.09}$ & $\mathbf{89.95}$ & $\mathbf{94.84}$ & $\mathbf{92.66}$ & $\mathbf{90.95}$ & $\mathbf{84.39}$ \\
\bottomrule
\end{tabular}
%\end{small}
\end{center}
\caption{Results with RoBERTa-base ($r=4$) on the GLUE benchmark datasets, comparing various assignment strategies for $\mathbf{A}_i$ and $\mathbf{B}_i$. We report Matthew's correlation for CoLA, Pearson correlation for STS-B, and accuracy for other datasets. Best results for each dataset are highlighted in \textbf{bold}.}
\label{tab:init_gluedatasets}
\end{table}

To extend our method to rank-heterogeneous settings, the assignments for $\mathbf{A}_i$ and $\mathbf{B}_i$ must also accommodate rank heterogeneity. Further investigation is required to develop an optimal assignment strategy that supports this.

\begin{figure}[ht]
    \centering
        \centering
        \includegraphics[width=0.6\textwidth]{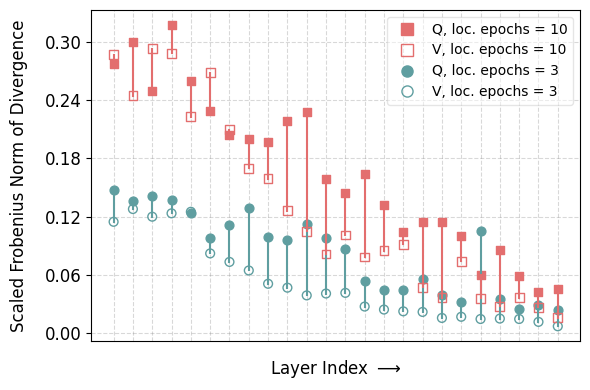}
        \caption{Scaled Frobenius norm of divergence/deviation of updates with conventional federated aggregation (FedAvg)  versus ideal LoRA updates, computed after the first aggregation step. We plot for query (Q) and value (V) matrices across model layers. Results are shown for local epochs $=\{\textcolor[HTML]{5F9EA0}{\textbf{3}}, \textcolor[HTML]{E36E6E}{\textbf{10}}\}$. (Dataset: MRPC, model: RoBERTa-large, $r=1$).}
        \label{fig:divergance-main}
\end{figure}

\textbf{Scaled Frobenius Norm of Divergence/Deviation.}
We now study the deviations in updates from federated averaging (FedAvg) relative to ideal updates and analyze the findings.
To quantify this deviation, we measure the scaled Frobenius norm of the divergence between the updates produced by FedAvg and the ideal LoRA updates, revealing several notable patterns. In Figure \ref{fig:divergance-main}, we plot this divergence for the query (Q) and value (V) matrices across model layers, computed after the first aggregation step for local epochs $=\{\textcolor[HTML]{5F9EA0}{\textbf{3}}, \textcolor[HTML]{E36E6E}{\textbf{10}}\}$. We observe that (1) the deviations decrease as the model depth increases, (2) the deviation grows with a higher number of local epochs, and (3) the deviation is more pronounced in the query (Q) matrices compared to the value (V) matrices.
% The following observations can be made:
% \begin{itemize}
%     \item The divergence decreases as the model depth increases.
%     \item The divergence increases with the number of local epochs.
%     \item The divergence is more pronounced in the query (Q) matrices compared to the value (V) matrices.
% \end{itemize}
These trends hold consistently across various datasets and settings, as shown by additional plots in Appendix \ref{app:div-1} (see Figures \ref{fig:div-app-1} and \ref{fig:div-app-2}).

\begin{figure}[ht]
    \centering
    % First subfigure
    \begin{subfigure}{0.49\textwidth}
        \centering
        \includegraphics[width=\textwidth]{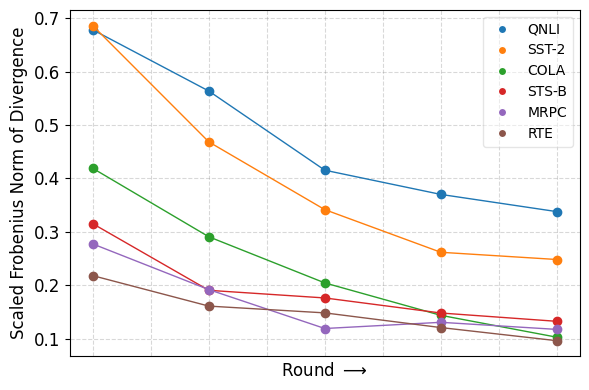}
        \caption{Query matrices of first layer}
        \label{fig:rounds-a0-main}
    \end{subfigure}
    \hfill
    % Second subfigure
    \begin{subfigure}{0.49\textwidth}
        \centering
        \includegraphics[width=\textwidth]{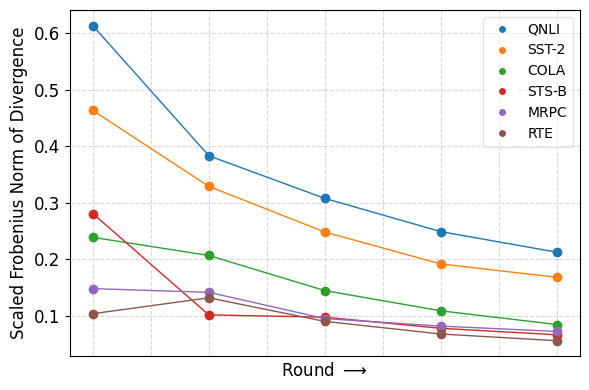}
        \caption{Avg. of query and value matrices across all layers}
        \label{fig:rounds-agg-main}
    \end{subfigure}
    \caption{Scaled Frobenius norm of divergence/deviation of updates with conventional federated aggregation (FedAvg) versus ideal LoRA updates, computed across multiple aggregation rounds for various datasets. We present results for (a) query matrices from the first layer, and (b) the average of query and value matrices across all layers. (Model: RoBERTa-large, $r=1$, local epochs $=10$).}
    \label{fig:rounds-main}
\end{figure}

Next, we examine how this deviation evolves across multiple rounds of federated aggregation. We plot the scaled Frobenius norm of the deviation between FedAvg and ideal LoRA updates over several aggregation rounds for different datasets, focusing on (a) the query matrices of the first layer, and (b) the average of the query and value matrices across all layers, as presented in Figure \ref{fig:rounds-main}. We observe that the deviation consistently decreases as the number of aggregation rounds increases, both for the first-layer query matrix and for the average of the query and value matrices across all layers.
These findings are further supported by detailed plots across multiple datasets and settings, as shown in Appendix \ref{app:div-2} (see Figures \ref{fig:rounds-app-1}, \ref{fig:rounds-app-2}, \ref{fig:rounds-app-3}, and \ref{fig:rounds-app-4}).

\textbf{Communication Costs.}
As discussed in Section \ref{sec:method}, FedEx-LoRA transmits a higher-rank update matrix (rank = $k \cdot r$) along with the low-rank adapters, which raises concerns about potential communication overhead. Table \ref{tab:comm-comparison} compares the communication costs of FFA-LoRA, FedIT, and full federated fine-tuning (FT), compared to FedEx-LoRA, for RoBERTa-base, RoBERTa-large, and GPT-2 models with rank $r=4$ over $5$ communication rounds. FedEx-LoRA incurs only a marginal increase in communication overhead relative to FedIT and FFA-LoRA, while FFA-LoRA has the lowest cost due to its reduced number of trainable parameters. FedEx-LoRA still maintains a substantially lower communication cost compared to federated full FT.

The practical impact of communication overhead is reduced by two factors: \textbf{(1)} the initial transmission of full model weights dominates communication costs, and \textbf{(2)} in NLU tasks, most communicated parameters come from the classification head, which requires training regardless of the aggregation method. Therefore, communication cost differences between FedEx-LoRA, FedIT, and FFA-LoRA are minimal in practice. Despite this marginal overhead, FedEx-LoRA consistently outperforms other federated LoRA approaches, making it an effective choice for federated fine-tuning.

\begin{table}[ht]
    \centering
    \small
    \begin{tabular}{lcccc}
        \toprule
        \textbf{Model} & \textbf{Federated Full FT} & \textbf{FedEx-LoRA} & \textbf{FedIT} & \textbf{FFA-LoRA} \\
        \midrule
        RoBERTa-base & $7.032$ & $1$ & $0.979$ & $0.972$ \\
        RoBERTa-large & $10.396$ & $1$ & $0.984$ & $0.979$ \\
        GPT-2 & $9.475$ & $1$ & $0.917$ & $0.886$ \\
        \bottomrule
    \end{tabular}
    \caption{Ratio of \# of parameters communicated in federated LoRA variants and federated full FT to FedEx-LoRA. All results are reported with rank $r = 4$ and across $5$ communication rounds.}
    \label{tab:comm-comparison}
\end{table}

\section{Conclusion}

In our work, we identified limitations in state-of-the-art federated fine-tuning methods that struggle with inexact aggregation. We proposed a novel method, FedEx-LoRA, which appends the residual error matrix to the frozen pretrained matrix, while maintaining minimal communication and computational overhead. The strength of our approach lies in its simplicity and broad applicability. Extensive experiments demonstrate that FedEx-LoRA consistently outperforms other federated LoRA methods across various datasets and settings. Our analyses reveal that deviations in updates from federated averaging compared to the ideal solution are significant and exhibit notable patterns.

Testing in privacy-preserving scenarios is a natural extension of our work. FFA-LoRA \citep{sun2024improving} demonstrated that noise in differential privacy leads to greater deviations from ideal updates. Given that our method achieves exact aggregation and outperforms FFA-LoRA in non-private settings, we anticipate similar success in privacy-sensitive applications. Our approach can be readily adapted for fine-tuning other models like Vision Transformers (ViTs) and Vision-Language models (VLMs).
\section*{Acknowledgements and Discolsure of Funding}

This work was supported by Mohamed bin Zayed University of Artificial Intelligence (MBZUAI) and partially funded by the ADIA Lab Fellowship.

\bibliography{neurips_2024}
\bibliographystyle{neurips_2024}

\clearpage

\appendix
\section{Dataset Details} \label{app:datasets}

\textsc{\textbf{CommonSense170K}} is a dataset combining eight commonsense reasoning datasets \citep{cr-dataset}, as detailed below: 

\begin{enumerate}
    \item \textbf{WinoGrande} \citep{sakaguchi2021winogrande} involves filling in blanks with binary choices based on sentences that demand commonsense reasoning.
    \item \textbf{HellaSwag} \citep{zellers2019hellaswag} asks the model to predict the most plausible continuation of a given context by selecting the correct ending from several options.
    \item \textbf{ARC Challenge} or \textbf{ARC-c} \citep{clark2018think} consists of multiple-choice science questions designed to challenge models with more complex reasoning, making them harder for methods that rely solely on co-occurrence patterns.
    \item \textbf{PIQA} \citep{bisk2020piqa} tests physical commonsense reasoning, where the task is to choose the best action from a set of options in a hypothetical situation.
    \item \textbf{BoolQ} \citep{clark2019boolq} focuses on yes/no question answering from naturally occurring queries.
    \item \textbf{ARC Easy} or \textbf{ARC-e} \citep{clark2018think} consists of grade-school-level multiple-choice science questions, providing a simpler set of tasks for testing models' basic reasoning abilities.
    \item \textbf{OBQA} \citep{mihaylov2018can} contains open-book, knowledge-intensive QA tasks requiring multi-hop reasoning to answer questions that involve integrating information from multiple sources.
    \item \textbf{SIQA} \citep{sap2019socialiqa} focuses on understanding human actions and predicting their social consequences, evaluating models' social commonsense reasoning.
\end{enumerate}

\textbf{MetaMathQA} dataset \citep{metamathqa} generates mathematical questions by rephrasing them from various perspectives without introducing additional knowledge. We evaluate this dataset on two benchmarks: GSM8K \citep{gsm8k}, which includes grade-school math word problems that require multi-step reasoning, and MATH \citep{math}, which features challenging competition-level mathematics problems.

\textbf{GLUE Benchmark} is a diverse suite of tasks for evaluating natural language understanding capabilities. It includes datasets such as SST-2 for sentiment analysis \citep{socher2013recursive}, MRPC for paraphrase detection \citep{dolan2005automatically}, CoLA for linguistic acceptability \citep{warstadt-etal-2019-neural}, QNLI for inference \citep{rajpurkar2018know}, RTE for inference, and STS-B for semantic textual similarity \citep{cer2017semeval}. Due to its comprehensive coverage of NLU tasks, GLUE is widely used to assess models like RoBERTa. Each dataset is released under its own license.

The \textbf{E2E NLG Challenge} \citep{novikova2017e2e} dataset is widely used to evaluate systems for natural language generation, particularly for data-to-text tasks. It contains around 42,000 training examples, with an additional 4,600 each for validation and testing, all from the restaurant domain. Each input table has multiple reference outputs, where each data point $(x, y)$ includes a sequence of slot-value pairs and its corresponding reference text in natural language. The dataset is made available under the Creative Commons BY-NC-SA 4.0 license.

\section{Hyperparameter Details} \label{app:hyperparams}

We conduct experiments on a single NVIDIA A100/A6000 GPU and report the average results from three independent runs. All models are trained using the AdamW optimizer \citep{loshchilov2019decoupledweightdecayregularization}. For the instruction tuning experiments, the hyperparameters and configurations for Mistral-7B, Gemma-2 9B, and Llama-3.2 3B are provided in Table \ref{tab:hyper_it}, following most of the settings from previous works \citep{cr-dataset}. The hyperparameter configurations for GPT-2 and RoBERTa-base/large are detailed in Table \ref{tab:hyper_gpt2_roberta}, with most settings following the original LoRA paper \citep{lora}, except for a learning rate sweep.

\begin{table}[ht]
\centering
\begin{tabular}{l|cc}
\hline
\toprule
 & \textbf{Mistral-7B / Gemma-2 9B} & \textbf{Llama-3.2 3B}\\
\midrule

Optimizer & \text{AdamW} & \text{AdamW} \\
Batch size & $1$ & $6$ \\
Max. Seq. Len & $512$ & $256$ \\
Grad Acc. Steps & $32$ & $24$ \\
Local Epochs & $1$ & $1$ \\
Rounds & $1$ & $1$ \\
Dropout & $0$ & $0$\\
Learning Rate & $5e-4$ & $5e-4$\\
LR Scheduler & Cosine & Linear\\
Warmup Ratio & $0.02$ & $0.02$\\
LoRA $\alpha$ & $16$ & $16$ \\
\bottomrule
\end{tabular}
\caption{Hyperparameter settings for Mistral-7B, Gemma-2 9B \& Llama-3.2 3B.}
\label{tab:hyper_it}
\end{table}

\begin{table}[ht]
\centering
\begin{tabular}{l|cc}
\hline
\toprule
 & \textbf{GPT-2} & \textbf{RoBERTa-base/large}\\
\midrule
&\multicolumn{2}{c}{Training} \\
\midrule
Optimizer & \text{AdamW} & \text{AdamW} \\
Weight Decay & $0.01$ & $0.01$\\
Dropout Prob & $0.1$ & $0.1$\\
Batch Size & $8$ & $128$ \\
Warmup Steps & $500$ & - \\
Warmup Ratio & - & $0.6$ \\
%LR Schedule & \text{Linear} & \text{Linear} \\
Label Smooth & $0.1$ & - \\
Max Seq. Len & $128$ & $512$ \\
Learning Rate & $2 \cdot 10^{-3}$ & $1 \cdot 10^{-3}$ \\
%Adaptation & $r_q=r_v=\{4, 1\}$ & $r_q=r_v=\{4, 1\}$\\
LoRA $\alpha$ & $32$ & $8$ \\
\midrule
&\multicolumn{2}{c}{Inference} \\
\midrule
Beam Size & $10$ & - \\
Length Penalty & $0.9$ & - \\
no repeat ngram size & $4$ & - \\
\bottomrule
\end{tabular}
\caption{Hyperparameter settings for GPT-2 and RoBERTa-base/large.}
\label{tab:hyper_gpt2_roberta}
\end{table}

\section {Effect of Varying Rank} \label{app:rank}
We evaluate FedEx-LoRA against other federated fine-tuning methods on the CoLA dataset using RoBERTa-base, by varying the rank of the low-rank adapters across $r = \{1, 2, 4, 8, 16, 32\}$, as presented in Table \ref{tab:fed_lora_ranks}. Across all rank configurations, FedEx-LoRA consistently outperforms competing federated LoRA variants. In agreement with prior studies \citep{lora, adalora}, increasing the rank does not always result in performance gains. For this task, we find that the optimal performance is achieved at $r=8$, beyond which further increases in rank yield diminishing returns.

\begin{table}[ht]
\centering
\small
\begin{tabular}{l|cccccc}
\hline
\toprule
{\bf Method} & {$\mathbf{r=1}$} & {$\mathbf{r=2}$} & {$\mathbf{r=4}$} & {$\mathbf{r=8}$} & {$\mathbf{r=16}$} & {$\mathbf{r=32}$} \\
\midrule
\rowcolor{gray!20} $\text{Centralized LoRA}$ & $62.13$ & $62.11$ & $64.31$ & $64.44$ & $64.32$ & $63.98$ \\
\rowcolor{cyan!10}$\text{FedIT}$ & $60.05$ & $60.32$ & $60.82$ & $62.09$ & $62.15$ & $61.98$ \\
\rowcolor{cyan!10}$\text{FFA-LoRA}$ & $57.73$ & $57.78$ & $59.34$ & $57.82$ & $57.78$ & $58.24$ \\
\rowcolor{cyan!10}$\text{FedEx-LoRA}$ & $\mathbf{62.07}$ & $\mathbf{61.38}$ & $\mathbf{62.82}$ & $\mathbf{63.57}$ & $\mathbf{63.56}$ & $\mathbf{63.35}$ \\
\bottomrule
\end{tabular}
\caption{Matthew's correlation on CoLA across different ranks for various federated LoRA methods. \textbf{Centralized LoRA (in grey) sets the benchmark skyline} for its federated versions. Best results among federated methods (in blue) are highlighted in \textbf{bold} for each rank. (Model: RoBERTa-base, local epochs $= 3$).}
\label{tab:fed_lora_ranks}
\end{table}

\section{Additional Experiments for NLU} \label{app:nlu}
We present additonal results with the RoBERTa-base and RoBERTa-large models in Table \ref{tab:app-gluedatasets}, evaluated at ranks $r = \{4, 1\}$, with local epochs set to $10$. 
%These findings corroborate the results in Table \ref{tab:gluedatasets}, demonstrating that our FedEx-LoRA method consistently outperforms state-of-the-art federated fine-tuning approaches across all datasets and configurations.

\begin{table}[ht]
    \begin{center}
        
        \begin{minipage}{\textwidth}
            \centering
            \small
            \begin{tabular}{l|ccccccc}
                \toprule
                \multirow{2}{*}{\bf Method} & {\bf CoLA} & {\bf RTE} & {\bf MRPC} & {\bf SST-2} & {\bf QNLI} & {\bf STS-B} & {\bf All} \\
                ~ & {Mcc $\uparrow$} & {Acc $\uparrow$} & {Acc $\uparrow$} & {Acc $\uparrow$} & {Acc $\uparrow$} & {Corr $\uparrow$} & {Avg $\uparrow$} \\
                \midrule
                \rowcolor{gray!20} $\text{Centralized LoRA}_{r=4}$ & $64.31$ & $75.45$ & $87.99$ & $94.61$ & $92.75$ & $90.73$ & $84.31$ \\
                \rowcolor{cyan!10} $\text{FedIT}_{r=4}$ & $58.55$ & $70.75$ & $87.50$ & $94.36$ & $92.09$ & $90.58$ & $82.31$ \\
                \rowcolor{cyan!10}$\text{FFA-LoRA}_{r=4}$ & $57.52$ & $71.84$ & $86.76$ & $94.24$ & $91.27$ & $90.04$ & $81.95$ \\
                \rowcolor{cyan!10}$\text{FedEx-LoRA}_{r=4}$ & $\mathbf{61.32}$ & $\mathbf{75.81}$ & $\mathbf{87.75}$ & $\mathbf{94.57}$ & $\mathbf{92.64}$ & $\mathbf{90.62}$ & $\mathbf{83.79}$ \\
                \midrule
                \rowcolor{gray!20} $\text{Centralized LoRA}_{r=1}$ & $62.13$ & $74.67$ & $87.75$ & $94.61$ & $92.31$ & $90.83$ & $83.72$ \\
                \rowcolor{cyan!10}$\text{FedIT}_{r=1}$ & $60.05$ & $71.84$ & $88.79$ & $94.62$ & $92.23$ & $90.54$ & $83.01$ \\
                \rowcolor{cyan!10}$\text{FFA-LoRA}_{r=1}$ & $57.73$ & $71.18$ & $87.74$ & $93.69$ & $91.41$ & $90.18$ & $81.99$ \\
                \rowcolor{cyan!10}$\text{FedEx-LoRA}_{r=1}$ & $\mathbf{61.31}$ & $\mathbf{73.12}$ & $\mathbf{89.21}$ & $\mathbf{94.73}$ & $\mathbf{92.40}$ & $\mathbf{90.67}$ & $\mathbf{83.57}$ \\
                \bottomrule
            \end{tabular}
            \caption*{(a) Results with RoBERTa-base on the GLUE benchmark datasets}
        \end{minipage}

        \vspace{0.5cm} % Space between subtables

        \begin{minipage}{\textwidth}
            \centering
            \small
            \begin{tabular}{l|ccccccc}
                \toprule
                \multirow{2}{*}{\bf Method} & {\bf CoLA} & {\bf RTE} & {\bf MRPC} & {\bf SST-2} & {\bf QNLI} & {\bf STS-B} & {\bf All} \\
                ~ & {Mcc $\uparrow$} & {Acc $\uparrow$} & {F1 $\uparrow$} & {Acc $\uparrow$} & {Acc $\uparrow$} & {Corr $\uparrow$} & {Avg $\uparrow$} \\
                \midrule
                \rowcolor{gray!20} $\text{Centralized LoRA}_{r=4}$ & $66.03$ & $82.67$ & $88.84$ & $96.21$ & $94.58$ & $91.92$ & $86.71$ \\
                \rowcolor{cyan!10}$\text{FedIT}_{r=4}$ & $61.80$ & $77.83$ & $85.54$ & $95.83$ & $94.32$ & $91.70$ & $84.50$ \\
                \rowcolor{cyan!10}$\text{FFA-LoRA}_{r=4}$ & $60.16$ & $74.67$ & $84.31$ & $95.64$ & $94.29$ & $90.28$ & $83.23$ \\
                \rowcolor{cyan!10}$\text{FedEx-LoRA}_{r=4}$ & $\mathbf{62.60}$ & $\mathbf{79.19}$ & $\mathbf{86.03}$ & $\mathbf{96.10}$ & $\mathbf{94.74}$ & $\mathbf{91.91}$ & $\mathbf{85.10}$ \\
                \midrule
                \rowcolor{gray!20} $\text{Centralized LoRA}_{r=1}$ & $65.21$ & $83.39$ & $89.21$ & $96.10$ & $94.42$ & $92.12$ & $86.74$ \\
                \rowcolor{cyan!10}$\text{FedIT}_{r=1}$ & $61.06$ & $78.33$ & $88.48$ & $95.86$ & $94.25$ & $91.17$ & $84.85$ \\
                \rowcolor{cyan!10}$\text{FFA-LoRA}_{r=1}$ & $60.32$ & $72.45$ & $85.78$ & $95.52$ & $93.94$ & $91.25$ & $83.21$ \\
                \rowcolor{cyan!10}$\text{FedEx-LoRA}_{r=1}$ & $\mathbf{63.56}$ & $\mathbf{79.07}$ & $\mathbf{89.71}$ & $\mathbf{96.22}$ & $\mathbf{94.56}$ & $\mathbf{91.77}$ & $\mathbf{85.82}$ \\
                \bottomrule
            \end{tabular}
            \caption*{(b) Results with RoBERTa-large on the GLUE benchmark datasets}
        \end{minipage}

        \caption{Results with RoBERTa-base and Roberta-large on the GLUE benchmark datasets, comparing various federated LoRA methods at ranks $r= \{4,1\}$. There are $10$ local epochs before every aggregation round.}
        \label{tab:app-gluedatasets}
        
    \end{center}
\end{table}

\section{Additional Experiments for NLG} \label{app:nlg}

Table \ref{tab:gpt2_ft_results-app} presents additional experiments  of GPT-2 fine-tuned with ranks $r = \{4, 1\}$, with local epochs set to $5$. FedEx-LoRA consistently outperforms leading federated fine-tuning methods across all metrics and settings, consistent with the results presented in Table \ref{tab:gpt2_ft_results}.

\begin{table}[!h]
\centering
\small
\begin{tabular}{l|ccccc}
\hline
\toprule
\multirow{2}*{\bf Method} & \multicolumn{5}{c}{\textbf{E2E NLG Challenge}} \\
 & BLEU $\uparrow$ & NIST $\uparrow$ & MET $\uparrow$ & ROUGE-L $\uparrow$ & CIDEr $\uparrow$\\
%\rowcolor{cyan!10} Full Fine-Tuning *  & $124.44$ M & $68.20$ & $8.62$ & $46.20$ & $71.00$ & $2.47$  \\
\midrule
\rowcolor{gray!20} $\text{Centralized LoRA}_{\text{r}=4}$ & $68.91$ & $8.73$ & $46.78$ & $71.29$ & $2.47$  \\
\rowcolor{cyan!10}$\text{FedIT}_{\text{r}=4}$  & $67.61$ & $8.62$ & $46.45$ & $70.28$ & $2.43$  \\
\rowcolor{cyan!10} $\text{FFA-LoRA}_{\text{r}=4}$   & $67.21$ & $8.57$ & $46.05$ & $69.98$ & $2.41$  \\
\rowcolor{cyan!10}$\text{Exact-FedIT}_{\text{r}=4}$ & $\mathbf{68.49}$ & $\mathbf{8.72}$ & $\mathbf{46.76}$ & $\mathbf{70.71}$ & $\mathbf{2.48}$  \\
\midrule
\rowcolor{gray!20} $\text{Centralized LoRA}_{\text{r}=1}$  & $67.41$ & $8.68$ & $46.01$ & $69.51$ & $2.41$  \\
\rowcolor{cyan!10}$\text{FedIT}_{\text{r}=1}$ & $66.16$ & $8.56$ & $45.54$ & $68.25$ & $2.29$  \\
\rowcolor{cyan!10} $\text{FFA-LoRA}_{\text{r}=1}$   & $65.78$ & $8.49$ & $45.01$ & $67.82$ & $2.26$  \\
\rowcolor{cyan!10}$\text{Exact-FedIT}_{\text{r}=1}$  & $\mathbf{66.54}$ & $\mathbf{8.57}$ & $\mathbf{46.07}$ & $\mathbf{69.11}$ & $\mathbf{2.37}$  \\
\bottomrule
\end{tabular}
\caption{Results with GPT-2 on the E2E NLG Challenge, comparing various federated LoRA methods at ranks $r= \{4,1\}$. There are $5$ local epochs before every aggregation round.}
\label{tab:gpt2_ft_results-app}
\end{table}

\clearpage

\section{More Divergence/Deviation Plots}\label{app:divergence-plots}

\subsection{Deviation/Divergence Plots Across Layers} \label{app:div-1}

\begin{figure}[ht]
    \centering
    % First row: QNLI
    \begin{subfigure}{0.49\textwidth}
        \centering
        \includegraphics[width=\textwidth]{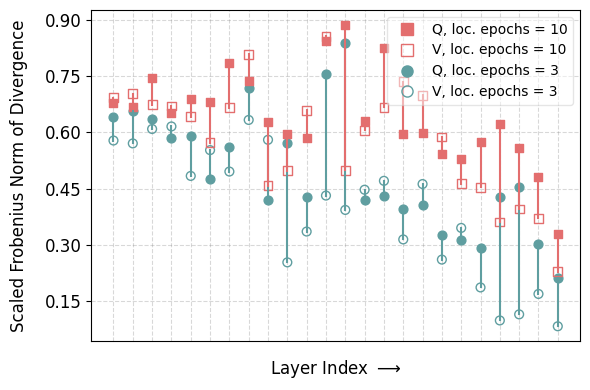}
        \caption{QNLI, $r=1$}
        \label{fig:4-bit-qnli}
    \end{subfigure}
    \hfill
    \begin{subfigure}{0.49\textwidth}
        \centering
        \includegraphics[width=\textwidth]{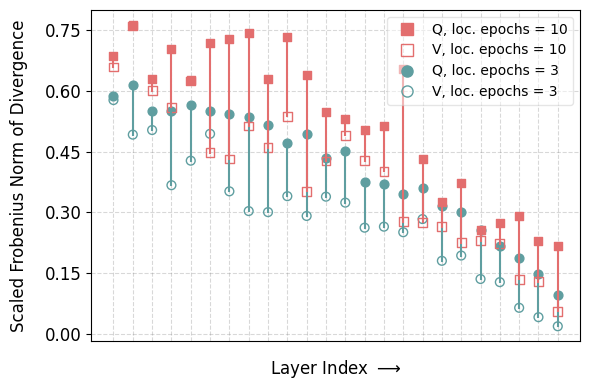}
        \caption{SST-2, $r=1$}
        \label{fig:3-bit-qnli}
    \end{subfigure}
    
    % Second row: SST2
    \begin{subfigure}{0.49\textwidth}
        \centering
        \includegraphics[width=\textwidth]{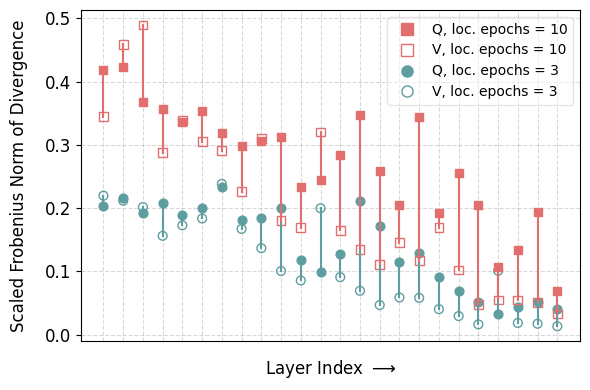}
        \caption{CoLA, $r=1$}
        \label{fig:4-bit-sst2}
    \end{subfigure}
    \hfill
    \begin{subfigure}{0.49\textwidth}
        \centering
        \includegraphics[width=\textwidth]{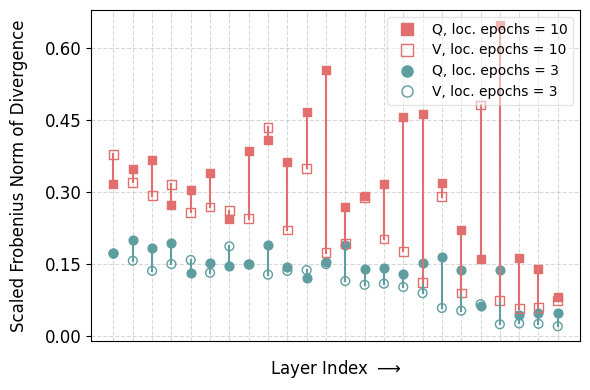}
        \caption{STS-B, $r=1$}
        \label{fig:3-bit-sst2}
    \end{subfigure}
    
    % Third row: COLA
    \begin{subfigure}{0.49\textwidth}
        \centering
        \includegraphics[width=\textwidth]{figures/mrpc_r1.png}
        \caption{MRPC, $r=1$}
        \label{fig:4-bit-cola}
    \end{subfigure}
    \hfill
    \begin{subfigure}{0.49\textwidth}
        \centering
        \includegraphics[width=\textwidth]{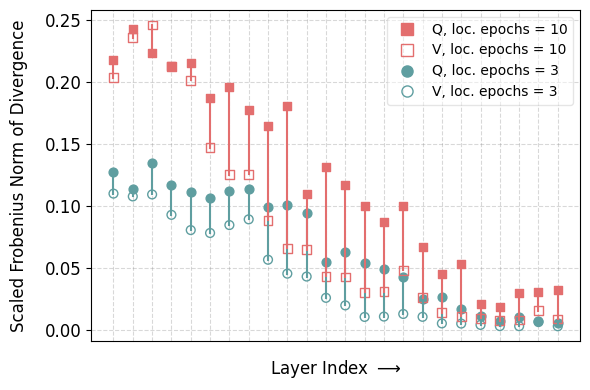}
        \caption{RTE, $r=1$}
        \label{fig:3-bit-cola}
    \end{subfigure}
    \caption{Scaled Frobenius norm of divergence/deviation of updates with conventional federated aggregation (FedAvg) versus ideal LoRA updates, computed after the first aggregation step. We plot for query (Q) and value (V) matrices across model layers, for multiple datasets. Results are shown for local epochs $=\{\textcolor[HTML]{5F9EA0}{\textbf{3}}, \textcolor[HTML]{E36E6E}{\textbf{10}}\}$. (Model: RoBERTa-large, $r=1$).}
    \label{fig:div-app-1}
\end{figure}

\begin{figure}[ht]
    \centering
    % First row: QNLI
    \begin{subfigure}{0.49\textwidth}
        \centering
        \includegraphics[width=\textwidth]{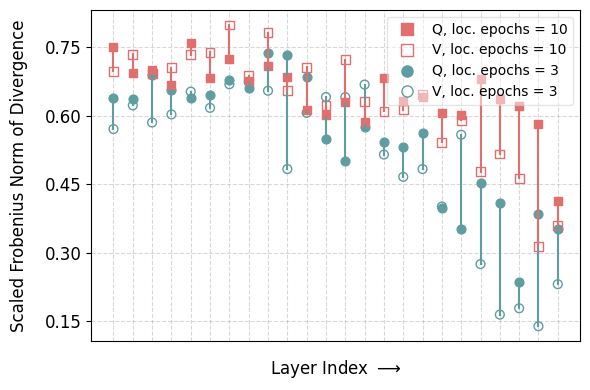}
        \caption{QNLI, $r=4$}
        \label{fig:4-bit-qnli}
    \end{subfigure}
    \hfill
    \begin{subfigure}{0.49\textwidth}
        \centering
        \includegraphics[width=\textwidth]{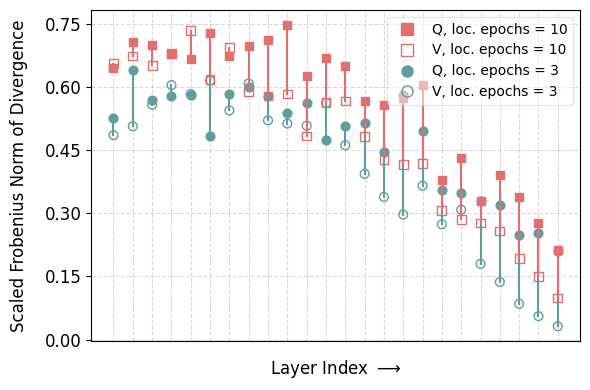}
        \caption{SST-2, $r=4$}
        \label{fig:3-bit-qnli}
    \end{subfigure}
    
    % Second row: SST2
    \begin{subfigure}{0.49\textwidth}
        \centering
        \includegraphics[width=\textwidth]{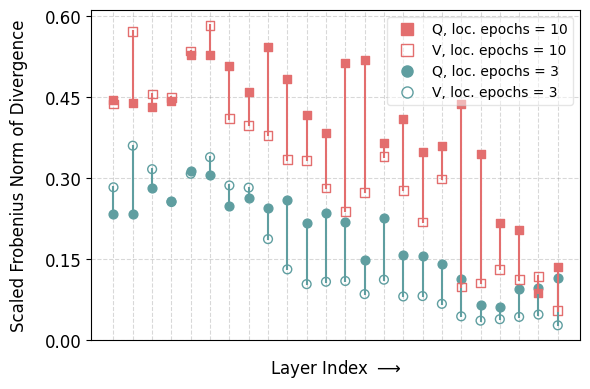}
        \caption{CoLA, $r=4$}
        \label{fig:4-bit-sst2}
    \end{subfigure}
    \hfill
    \begin{subfigure}{0.49\textwidth}
        \centering
        \includegraphics[width=\textwidth]{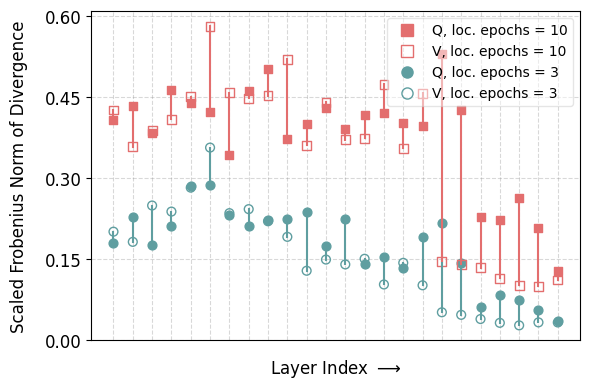}
        \caption{STS-B, $r=4$}
        \label{fig:3-bit-sst2}
    \end{subfigure}
    
    % Third row: COLA
    \begin{subfigure}{0.49\textwidth}
        \centering
        \includegraphics[width=\textwidth]{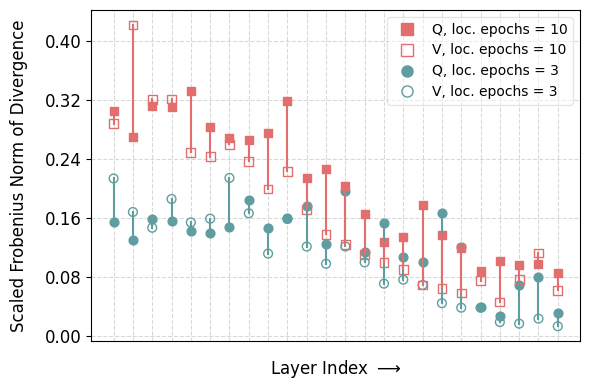}
        \caption{MRPC, $r=4$}
        \label{fig:4-bit-cola}
    \end{subfigure}
    \hfill
    \begin{subfigure}{0.49\textwidth}
        \centering
        \includegraphics[width=\textwidth]{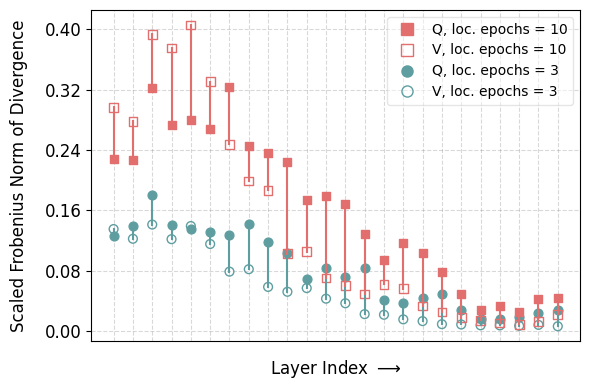}
        \caption{RTE, $r=4$}
        \label{fig:dddd}
    \end{subfigure}
    \caption{Scaled Frobenius norm of divergence/deviation of updates with conventional federated aggregation (FedAvg) versus ideal LoRA updates, computed after the first aggregation step. We plot for query (Q) and value (V) matrices across model layers, for multiple datasets. Results are shown for local epochs $=\{\textcolor[HTML]{5F9EA0}{\textbf{3}}, \textcolor[HTML]{E36E6E}{\textbf{10}}\}$. (Model: RoBERTa-large, $r=4$).}
    \label{fig:div-app-2}
\end{figure}

As discussed in Section \ref{sec:analysis}, we further quantify the deviation of conventional federated aggregation (FedAvg) from ideal updates by measuring the scaled Frobenius norm of the divergence the updates produced by FedAvg and the ideal LoRA updates. We present additional plots of this divergence for the query (Q) and value (V) matrices across model layers, computed after the first aggregation step for local epochs $=\{\textcolor[HTML]{5F9EA0}{\textbf{3}}, \textcolor[HTML]{E36E6E}{\textbf{10}}\}$ across multiple datasets, in Figures \ref{fig:div-app-1} and \ref{fig:div-app-2}. Figure \ref{fig:div-app-1} shows results for rank $r=1$, while Figure \ref{fig:div-app-2} presents results for rank $r=4$.

% \begin{figure}[ht]
%     \centering
%     % First subfigure
%     \begin{subfigure}{0.49\textwidth}
%         \centering
%         \includegraphics[width=\textwidth]{figures/rounds_a0_l10_r1.png}
%         \caption{Query matrices of first layer, local epochs $=3$}
%         \label{fig:4-bit-fault-aware}
%     \end{subfigure}
%     \hfill
%     % Second subfigure
%     \begin{subfigure}{0.49\textwidth}
%         \centering
%         \includegraphics[width=\textwidth]{figures/rounds_avg_l10_r1.png}
%         \caption{3}
%         \label{fig:3-bit-fault-aware}
%     \end{subfigure}

%         \begin{subfigure}{0.49\textwidth}
%         \centering
%         \includegraphics[width=\textwidth]{figures/rounds_a0_l3_r1.png}
%         \caption{4}
%         \label{fig:4-bit-fault-aware}
%     \end{subfigure}
%     \hfill
%     % Second subfigure
%     \begin{subfigure}{0.49\textwidth}
%         \centering
%         \includegraphics[width=\textwidth]{figures/rounds_avg_l3_r1.png}
%         \caption{3}
%         \label{fig:3-bit-fault-aware}
%     \end{subfigure}
    
%     \caption{Scaled Frobenius norm of divergence/deviation of updates with conventional federated aggregation versus ideal LoRA updates, computed across multiple aggregation rounds for various datasets. We present results for (a, c) query matrices from the first layer, and (b, d) the average of query and value matrices across all layers. (Model: RoBERTa-large, $r=1$)}
%     \label{fig:mainfigure}
% \end{figure}
\clearpage

\subsection{Deviation/Divergence Plots Across Rounds} \label{app:div-2}

\begin{figure}[H]
    \centering
    % First subfigure
    \begin{subfigure}{0.49\textwidth}
        \centering
        \includegraphics[width=\textwidth]{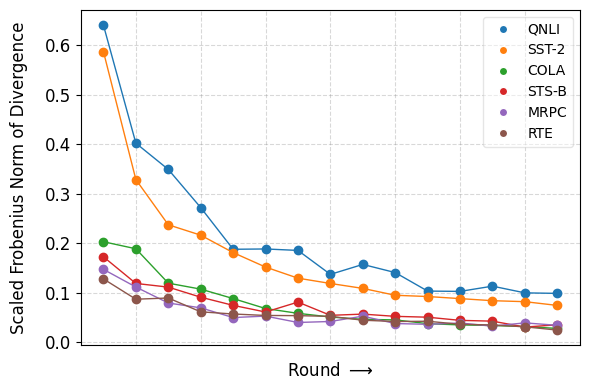}
        \caption{Query matrices of first layer}
        \label{fig:13}
    \end{subfigure}
    \hfill
    % Second subfigure
    \begin{subfigure}{0.49\textwidth}
        \centering
        \includegraphics[width=\textwidth]{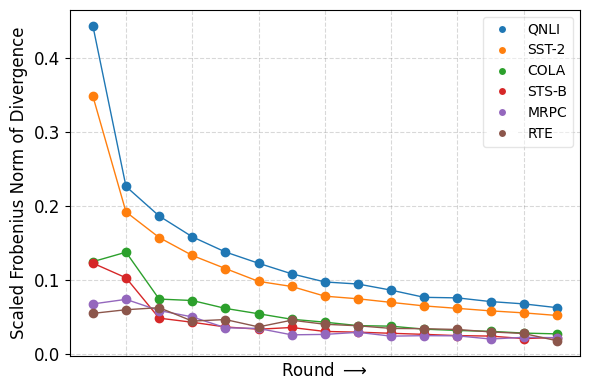}
        \caption{Avg. of query and value matrices across all layers}
        \label{fig:3-bit-fault-aware}
    \end{subfigure}
    \caption{Scaled Frobenius norm of divergence/deviation of updates with conventional federated aggregation (FedAvg) versus ideal LoRA updates, computed across multiple aggregation rounds for various datasets. We present results for (a) query matrices from the first layer, and (b) the average of query and value matrices across all layers. (Model: RoBERTa-large, $r=1$, local epochs = $3$)}
    \label{fig:rounds-app-1}
\end{figure}

\begin{figure}[H]
    \centering
    % First subfigure
    \begin{subfigure}{0.49\textwidth}
        \centering
        \includegraphics[width=\textwidth]{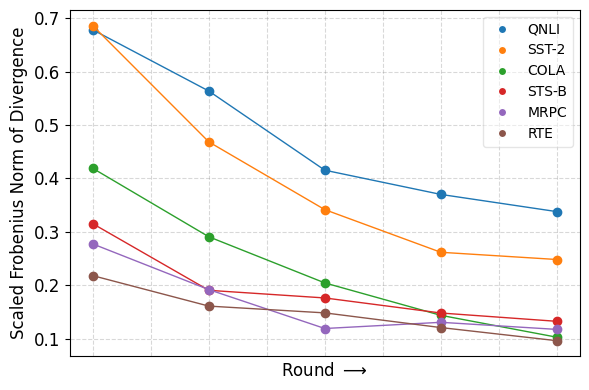}
        \caption{Query matrices of first layer}
        \label{fig:4-bit-fault-aware}
    \end{subfigure}
    \hfill
    % Second subfigure
    \begin{subfigure}{0.49\textwidth}
        \centering
        \includegraphics[width=\textwidth]{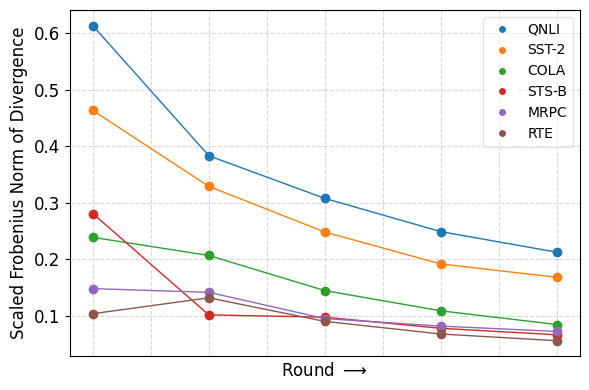}
        \caption{Avg. of query and value matrices across all layers}
        \label{fig:3-bit-fault-aware}
    \end{subfigure}
    \caption{Scaled Frobenius norm of divergence/deviation of updates with conventional federated aggregation (FedAvg) versus ideal LoRA updates, computed across multiple aggregation rounds for various datasets. We present results for (a) query matrices from the first layer, and (b) the average of query and value matrices across all layers. (Model: RoBERTa-large, $r=1$, local epochs = $10$)}
    \label{fig:rounds-app-2}
\end{figure}

\begin{figure}[H]
    \centering
    % First subfigure
    \begin{subfigure}{0.49\textwidth}
        \centering
        \includegraphics[width=\textwidth]{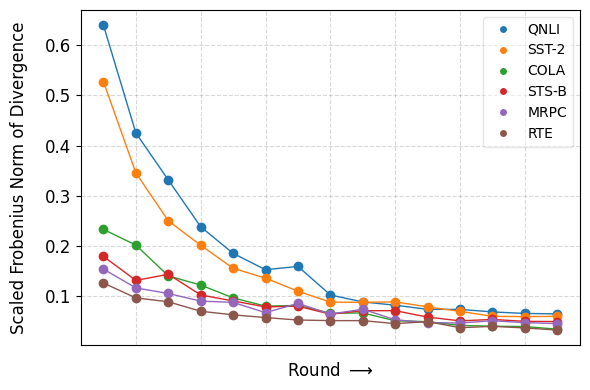}
        \caption{Query matrices of first layer}
        \label{fig:4-bit-fault-aware}
    \end{subfigure}
    \hfill
    % Second subfigure
    \begin{subfigure}{0.49\textwidth}
        \centering
        \includegraphics[width=\textwidth]{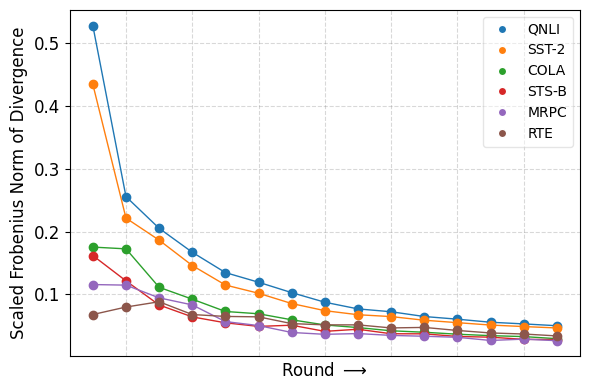}
        \caption{Avg. of query and value matrices across all layers}
        \label{fig:3-bit-fault-aware}
    \end{subfigure}
    \caption{Scaled Frobenius norm of divergence/deviation of updates with conventional federated aggregation (FedAvg) versus ideal LoRA updates, computed across multiple aggregation rounds for various datasets. We present results for (a) query matrices from the first layer, and (b) the average of query and value matrices across all layers. (Model: RoBERTa-large, $r=4$, local epochs = $3$)}
    \label{fig:rounds-app-3}
\end{figure}

\begin{figure}[H]
    \centering
    % First subfigure
    \begin{subfigure}{0.49\textwidth}
        \centering
        \includegraphics[width=\textwidth]{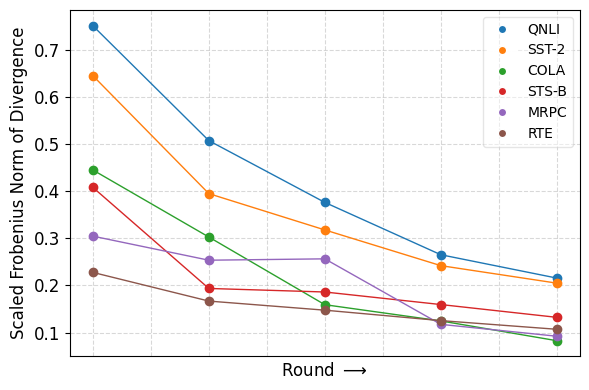}
        \caption{Query matrices of first layer}
        \label{fig:4-bit-fault-aware}
    \end{subfigure}
    \hfill
    % Second subfigure
    \begin{subfigure}{0.49\textwidth}
        \centering
        \includegraphics[width=\textwidth]{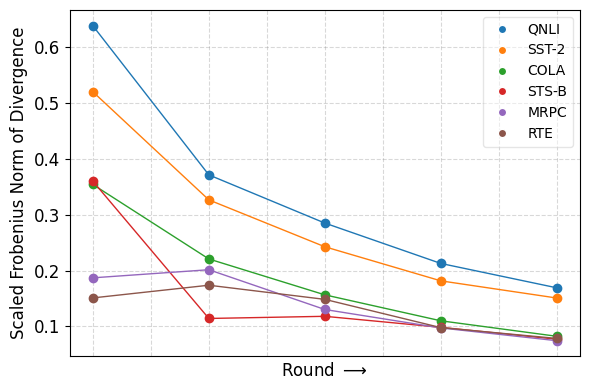}
        \caption{Avg. of query and value matrices across all layers}
        \label{fig:3-bit-fault-aware}
    \end{subfigure}
    \caption{Scaled Frobenius norm of divergence/deviation of updates with conventional federated aggregation (FedAvg) versus ideal LoRA updates, computed across multiple aggregation rounds for various datasets. We present results for (a) query matrices from the first layer, and (b) the average of query and value matrices across all layers. (Model: RoBERTa-large, $r=4$, local epochs = $10$)}
    \label{fig:rounds-app-4}
\end{figure}

We now examine how the deviation evolves across multiple rounds of federated aggregation. We plot the scaled Frobenius norm of the deviation between FedAvg and ideal LoRA updates over several aggregation rounds for different datasets, focusing on (a) the query matrices of the first layer and (b) the average of the query and value matrices across all layers. This is presented in Figures \ref{fig:rounds-app-1}, \ref{fig:rounds-app-2}, \ref{fig:rounds-app-3}, and \ref{fig:rounds-app-4}. We include results for ranks $r = \{1,4\}$ and local epochs $=\{3, 10\}$.

 \clearpage
 
\end{document}